\documentclass[ALICE,manyauthors]{cernphprep}
\usepackage[comma,square,numbers,sort&compress]{natbib}
\usepackage{hyperref}
\usepackage{lineno}
\usepackage{xspace}
\usepackage{xcolor}
\usepackage{booktabs}
\usepackage{multicol}
\usepackage{multirow}
\usepackage{rotating}
\usepackage[T1]{fontenc}
\usepackage{orcidlink}

\begin{document}
%

\newcommand{\pp}           {pp\xspace}
\newcommand{\ppbar}        {\mbox{$\mathrm {p\overline{p}}$}\xspace}
\newcommand{\XeXe}         {\mbox{Xe--Xe}\xspace}
\newcommand{\PbPb}         {\mbox{Pb--Pb}\xspace}
\newcommand{\pA}           {\mbox{pA}\xspace}
\newcommand{\pPb}          {\mbox{p--Pb}\xspace}
\newcommand{\AuAu}         {\mbox{Au--Au}\xspace}
\newcommand{\dAu}          {\mbox{d--Au}\xspace}
\newcommand{\Pbp}          {\mbox{Pb--p}\xspace}

\newcommand{\s}            {\ensuremath{\sqrt{s}}\xspace}
\newcommand{\snn}          {\ensuremath{\sqrt{s_{\mathrm{NN}}}}\xspace}
\newcommand{\pt}           {\ensuremath{p_{\rm T}}\xspace}
\newcommand{\meanpt}       {$\langle p_{\mathrm{T}}\rangle$\xspace}
\newcommand{\ycms}         {\ensuremath{y_{\rm CMS}}\xspace}
\newcommand{\ylab}         {\ensuremath{y_{\rm lab}}\xspace}
\newcommand{\etalab}       {\ensuremath{\eta_{\rm lab}}\xspace}
\newcommand{\etarange}[1]  {\mbox{$\left | \eta \right |~<~#1$}}
\newcommand{\yrange}[1]    {\mbox{$\left | y \right |~<~#1$}}
\newcommand{\dndy}         {\ensuremath{\mathrm{d}N_\mathrm{ch}/\mathrm{d}y}\xspace}
\newcommand{\dndeta}       {\ensuremath{\mathrm{d}N_\mathrm{ch}/\mathrm{d}\eta}\xspace}
\newcommand{\dndetalab}       {\ensuremath{\mathrm{d}N_\mathrm{ch}/\mathrm{d}\eta_{\rm lab}}\xspace}
\newcommand{\dndetaphoton} {\ensuremath{\mathrm{d}N_\mathrm{\gamma}/\mathrm{d}\eta}\xspace}
\newcommand{\dndetaphotonlab} {\ensuremath{\mathrm{d}N_\mathrm{\gamma}/\mathrm{d}\eta_{\rm lab}}\xspace}
\newcommand{\avdndeta}     {\ensuremath{\langle\dndeta\rangle}\xspace}
\newcommand{\dNdy}         {\ensuremath{\mathrm{d}N_\mathrm{ch}/\mathrm{d}y}\xspace}
\newcommand{\Npart}        {\ensuremath{N_\mathrm{part}}\xspace}
\newcommand{\Ncoll}        {\ensuremath{N_\mathrm{coll}}\xspace}
\newcommand{\dEdx}         {\ensuremath{\textrm{d}E/\textrm{d}x}\xspace}
\newcommand{\RpPb}         {\ensuremath{R_{\rm pPb}}\xspace}

\newcommand{\nineH}        {$\sqrt{s}~=~0.9$~Te\kern-.1emV\xspace}
\newcommand{\seven}        {$\sqrt{s}~=~7$~Te\kern-.1emV\xspace}
\newcommand{\twoH}         {$\sqrt{s}~=~0.2$~Te\kern-.1emV\xspace}
\newcommand{\twosevensix}  {$\sqrt{s}~=~2.76$~Te\kern-.1emV\xspace}
\newcommand{\five}         {$\sqrt{s}~=~5.02$~Te\kern-.1emV\xspace}
\newcommand{\twosevensixnn}{$\sqrt{s_{\mathrm{NN}}}~=~2.76$~Te\kern-.1emV\xspace}
\newcommand{\fivenn}       {$\sqrt{s_{\mathrm{NN}}}~=~5.02$~Te\kern-.1emV\xspace}
\newcommand{\LT}           {L{\'e}vy-Tsallis\xspace}
\newcommand{\GeVc}         {Ge\kern-.1emV/$c$\xspace}
\newcommand{\MeVc}         {Me\kern-.1emV/$c$\xspace}
\newcommand{\TeV}          {Te\kern-.1emV\xspace}
\newcommand{\GeV}          {Ge\kern-.1emV\xspace}
\newcommand{\MeV}          {Me\kern-.1emV\xspace}
\newcommand{\GeVmass}      {Ge\kern-.2emV/$c^2$\xspace}
\newcommand{\MeVmass}      {Me\kern-.2emV/$c^2$\xspace}
\newcommand{\lumi}         {\ensuremath{\mathcal{L}}\xspace}

\newcommand{\ITS}          {\rm{ITS}\xspace}
\newcommand{\TOF}          {\rm{TOF}\xspace}
\newcommand{\ZDC}          {\rm{ZDC}\xspace}
\newcommand{\ZDCs}         {\rm{ZDCs}\xspace}
\newcommand{\ZNA}          {\rm{ZNA}\xspace}
\newcommand{\ZNC}          {\rm{ZNC}\xspace}
\newcommand{\SPD}          {\rm{SPD}\xspace}
\newcommand{\SDD}          {\rm{SDD}\xspace}
\newcommand{\SSD}          {\rm{SSD}\xspace}
\newcommand{\TPC}          {\rm{TPC}\xspace}
\newcommand{\TRD}          {\rm{TRD}\xspace}
\newcommand{\VZERO}        {\rm{V0}\xspace}
\newcommand{\VZEROA}       {\rm{V0A}\xspace}
\newcommand{\VZEROC}       {\rm{V0C}\xspace}
\newcommand{\Vdecay} 	   {\ensuremath{V^{0}}\xspace}
\newcommand{\PMD}          {\rm{PMD}\xspace}
\newcommand{\ZPA}          {\rm{ZPA}\xspace}
\newcommand{\ZPC}          {\rm{ZPC}\xspace}

\newcommand{\ee}           {\ensuremath{e^{+}e^{-}}} 
\newcommand{\pip}          {\ensuremath{\pi^{+}}\xspace}
\newcommand{\pim}          {\ensuremath{\pi^{-}}\xspace}
\newcommand{\pizero}        {\ensuremath{\pi^{0}}\xspace}
\newcommand{\kap}          {\ensuremath{\rm{K}^{+}}\xspace}
\newcommand{\kam}          {\ensuremath{\rm{K}^{-}}\xspace}
\newcommand{\pbar}         {\ensuremath{\rm\overline{p}}\xspace}
\newcommand{\kzero}        {\ensuremath{{\rm K}^{0}_{\rm{S}}}\xspace}
\newcommand{\lmb}          {\ensuremath{\Lambda}\xspace}
\newcommand{\almb}         {\ensuremath{\overline{\Lambda}}\xspace}
\newcommand{\Om}           {\ensuremath{\Omega^-}\xspace}
\newcommand{\Mo}           {\ensuremath{\overline{\Omega}^+}\xspace}
\newcommand{\X}            {\ensuremath{\Xi^-}\xspace}
\newcommand{\Ix}           {\ensuremath{\overline{\Xi}^+}\xspace}
\newcommand{\Xis}          {\ensuremath{\Xi^{\pm}}\xspace}
\newcommand{\Oms}          {\ensuremath{\Omega^{\pm}}\xspace}
\newcommand{\degree}       {\ensuremath{^{\rm o}}\xspace}

\newcommand{\pyperu}       {\rm{PYTHIA~6~Perugia~2011}\xspace}
\newcommand{\pymonash}     {\rm{PYTHIA~8~Monash~2013}\xspace}
\newcommand{\phojet}       {\rm{PHOJET}\xspace}
\newcommand{\eposlhc}      {\rm{EPOS~LHC}\xspace}
\newcommand{\ampt}         {\rm{AMPT}\xspace}
\newcommand{\dpmjet}       {\textsc{Dpmjet}\xspace}
\newcommand{\hijing}       {\textsc{Hijing}\xspace}
\newcommand{\angantyr}     {PYTHIA 8/Angantyr\xspace}

\begin{titlepage}
\PHyear{2025}       
\PHnumber{023}      
\PHdate{17 February}  
  
  \title{Charged-particle multiplicity distributions over a wide pseudorapidity range
    in \pPb collisions at $\sqrt{\textbf{\textit{s}}_{\rm \textbf{NN}}}=\textbf{5.02}$ TeV}
  \ShortTitle{Charged-particle multiplicity in \pPb collisions}   
  
  \Collaboration{ALICE Collaboration\thanks{See Appendix~\ref{app:collab} for the list of collaboration members}}
  \ShortAuthor{ALICE Collaboration} 

  \begin{abstract}
    This paper presents the primary charged-particle multiplicity distributions in proton--lead collisions at a centre-of-mass energy per nucleon--nucleon collision of \fivenn. The distributions are reported for non-single diffractive collisions in different pseudorapidity ranges. The measurements are performed using the combined information from the Silicon Pixel Detector and the Forward Multiplicity Detector of ALICE. The multiplicity distributions are parametrised with a double negative binomial distribution function which provides satisfactory descriptions of the distributions for all the studied pseudorapidity intervals. The data are compared to models and analysed quantitatively, evaluating  the first four moments (mean, standard deviation, skewness, and kurtosis). The shape evolution of the measured multiplicity distributions is studied in terms of KNO variables and it is found that none of the considered models reproduces the measurements. This paper also reports on the average charged-particle multiplicity, normalised by the average number of participating nucleon pairs, as a function of the collision energy. The multiplicity results are then compared to measurements made in proton--proton and nucleus--nucleus collisions across a wide range of collision energies.
    
  \end{abstract}
\end{titlepage}

\setcounter{page}{2} 


\section{Introduction}
The multiplicity distribution of primary charged particles, P($N_{\rm ch}$), is one of the key observables that provides valuable insights into the particle production mechanisms in high-energy hadronic and nuclear collisions. The production of charged particles at current collider energies involves the interplay of perturbative and non-perturbative quantum chromodynamic (QCD) interactions and is sensitive to colliding particle species, centre-of-mass energy, and collision centrality. ALICE measurements of charged-particle multiplicities across different collision systems over a broad range of pseudorapidity allow us to perform comprehensive studies of particle production at Large Hadron Collider (LHC) energies~\cite{ALICE:ppMidChMultpaper900and2360,ALICE:ppMidChMultpaper7000,ALICE:ppMidChMultpaper900to8000,ALICE:ppFrdChMultpaper,ALICE:PbPbsatellite,ALICE:PbPb_CentFrdrapidity2760,ALICE:PbPb_CentFrdrapidity5020,ALICE:2022imr}.

Recent experimental findings in proton--lead (\pPb) collisions have shown characteristics of collectivity and strangeness enhancement that are typically attributed in heavy-ion collisions to the creation of a quark--gluon plasma (QGP)~\cite{ALICE:RidgepPb,ATLAS:2012cix,CMS:2012qk,ATLAS:2014qaj,ALICE:ReviewPaper}. The origin of these phenomena is not yet fully understood, and it is crucial to investigate and understand the global properties of the system formed in \pPb collisions, which makes the measurement of multiplicity distributions important. Moreover, the study of \pPb collisions aids in understanding cold nuclear matter effects~\cite{Qiu:2004da,Wang:2001ifa} on the final-state particle production.

Following earlier ALICE results in proton--proton (\pp) collisions~\cite{ALICE:ppFrdChMultpaper}, this paper presents, for the first time in \pPb collisions at \fivenn, a comprehensive set of measurements of P($N_{\rm ch}$) for the full phase space ($-3.4<\etalab<5.0$) and for a set of symmetric pseudorapidity ranges: $|\etalab|<2.4$, $|\etalab|<3.0$, and $|\etalab|<3.4$. We also study the charged-particle production on both the p-fragmentation and the Pb-fragmentation sides in \pPb collisions, covering pseudorapidity ranges: $-3.4<\etalab<-1.0$ and $2.0<\etalab<5.0$, respectively. The results are compared to model calculations from HIJING (v1.36)~\cite{hijing}, DPMJET (v3.0-5)~\cite{dpmjet}, PYTHIA 8.308/Angantyr~\cite{Angantyr}, and QCD saturation-based IP-Glasma~\cite{Schenke:2012wb,Schenke:2012hg}. From the multiplicity distributions, we calculate the mean ($\langle N_{\rm ch}\rangle$), standard deviation ($\sigma$), skewness ($S$), and kurtosis ($\kappa$) and compare them to the same moments evaluated from the considered models. This approach allows for a quantitative comparison of the performance of these models and provides input for their improved tuning to accurately simulate the underlying physics processes involved in particle production. This paper also reports a description of multiplicity distributions in terms of a double negative binomial distribution (NBD) function.

This article is organised as follows: Section~\ref{exsetup} describes the experimental conditions, data sample considered in the analysis, the selection of collisions, and the reconstruction of charged particles. Section~\ref{correction} explains the correction procedure applied to the data. The estimates of systematic uncertainties from various sources are discussed in Sec.~\ref{SysUncEst}. Section~\ref{results} presents the results of this analysis, and, finally, the conclusions are summarised in Sec.~\ref{summary}.

\section{Experimental details}
\label{exsetup}
The full description of the ALICE detectors and their performance can be found in dedicated publications~\cite{ALICE:Exp,ALICE:performance,ALICE:ReviewPaper}. The ALICE reference frame is defined with the $z$ axis directed along the beam line and the nominal interaction point (IP) at $z=0$. This analysis uses the data collected by ALICE in 2013 during the \pPb collision run of the LHC. In these collisions, a proton beam with an energy of 4\,\TeV circulated towards the negative $z$ direction ($\etalab<0$), while lead ions with an energy of 1.58\,\TeV per nucleon circulated in the opposite direction ($\etalab>0$). This configuration resulted in collisions at \fivenn in the nucleon--nucleon centre-of-mass frame which is shifted in rapidity by $\Delta y=0.465$ in the direction of the proton beam. In the following, the variable \etalab represents the pseudorapidity in the laboratory reference frame. The sub-detectors used in this analysis are briefly described below.

The V0 detector~\cite{ALICE:FrdDettdr,ALICE:V0performance} is made of two arrays of 32 scintillators: V0A, positioned at $z=330$ cm and covering the pseudorapidity interval $2.8<\etalab<5.1$, and V0C, at $z=-90$ cm and covering \linebreak $-3.7<\etalab<-1.7$. Both the amplitude and the time of the signals produced by charged particles that hit each scintillator are recorded. The V0 detector is used for minimum-bias trigger selection and background rejection in this analysis. 

The Silicon Pixel Detector (SPD) consists of the two innermost cylindrical layers of the ALICE Inner Tracking System (ITS)~\cite{ALICE:performance,ALICE:ITStdr} surrounding the central beryllium beam pipe. The SPD covers the pseudorapidity ranges $|\etalab|<2$ and $|\etalab|<1.4$ with full azimuthal coverage for the inner and outer layers, respectively. In this analysis, the SPD is used to determine the position of the interaction vertex and to estimate the charged-particle multiplicity around midrapidity ($|\etalab|<2$).

The Forward Multiplicity Detector (FMD)~\cite{ALICE:FrdDettdr,ALICE:PbPb_CentFrdrapidity2760,ALICE:ppFrdChMultpaper} is a silicon strip detector composed of three sub-detectors placed at $z = 320$\,cm (FMD1), 79\,cm (FMD2), and $-$69\,cm (FMD3). The FMD has full azimuthal coverage in the pseudorapidity ranges $-3.4<\etalab<-1.7$ (FMD3) and $1.7<\etalab<5.0$ (FMD1 and FMD2), and these extend the charged-particle detection acceptance beyond the reach of the central detectors in ALICE.

A sample of non-single diffractive (NSD) collisions is selected using a minimum-bias (MB) trigger condition, which requires a coincidence between V0A and V0C time signals. The standard ALICE collision selection criteria~\cite{ALICE:pPbMidChdNdEtapaper5020MB} is used in this analysis, which includes: rejection of background collisions such as beam--gas or beam--halo interactions that occur outside the interaction region, exclusion of pile-up collisions, and selection of the reconstructed primary vertex position ($z_{\mathrm{vtx}}$) along $z$ axis. However, in this analysis, the position of $z_{\mathrm{vtx}}$ is further restricted to be within $\pm$4\,cm from the nominal IP to minimise the acceptance gaps in the pseudorapidity coverage of the SPD and FMD~\cite{ALICE:ppFrdChMultpaper}. After applying all selection criteria, approximately 9 million \pPb collisions are considered in this analysis.

The measurements of multiplicity at mid and forward rapidity are provided by the SPD and FMD, respectively. This analysis is focused on primary charged-particle measurements. Primary charged particles are defined as charged particles with a mean proper lifetime $\tau$ larger than 1\,cm/$c$, which are either a) produced directly in the collision, or b) from decays of particles with $\tau$ smaller than 1\,cm/$c$, excluding particles produced in interactions with material~\cite{ALICE:ChPrDefinition}. In the midrapidity region ($|\etalab|<2$), charged particles can deposit energy and produce signals in more than one pixel of the SPD. The offline reconstruction combines such adjacent pixel signals into a single cluster. The clusters from the two layers of SPD, together with the primary vertex, are combined to form tracklets. The charged-particle multiplicity is then determined by counting the number of tracklets~\cite{ALICE:MBMidrapidity2760}. In the forward regions ($-3.4<\etalab<-1.7$ and $1.7<\etalab<5.0$), the FMD records the energy deposited by charged particles that traverse each silicon strip. The number of charged particles per strip is then calculated using a statistical approach as described in Ref.~\cite{ALICE:PbPbsatellite}. When there is an overlap in the acceptance ($1.7<|\etalab|<2$) of the SPD and FMD, the multiplicity is determined by averaging the two measurements.

\begin{figure}
  \centering
  \includegraphics[scale=0.7]{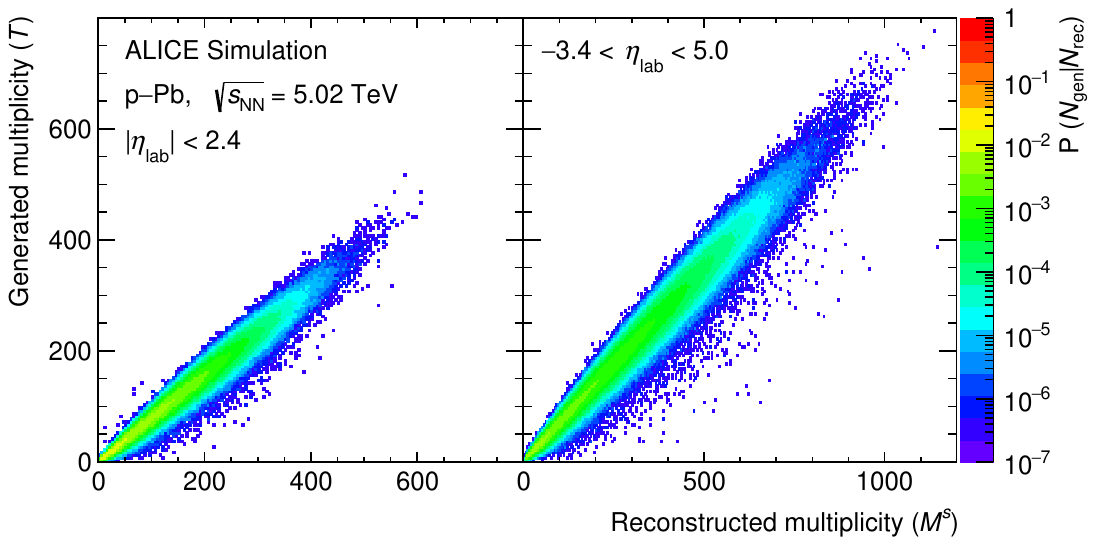}
  \caption{Graphical representation of the detector response matrices obtained with the HIJING event generator for two pseudorapidity coverages: $|\etalab|<2.4$ (left) and $-3.4<\etalab<5.0$ (right) in \pPb collisions at \fivenn.}
  \label{response-matrix}
\end{figure}

\section{Correction procedure}
\label{correction}
As reported earlier, the main challenge in measuring the charged-particle multiplicity at forward rapidity is the significant background of secondary particles produced in interactions with the beam pipe and the material that exists in front of the FMD~\cite{ALICE:ppFrdChMultpaper,ALICE:PbPbsatellite,ALICE:PbPb_CentFrdrapidity2760}. There are also other instrumental effects, such as detector acceptance and collision selection inefficiencies. A set of correction techniques is considered to account for these effects.

The main ingredients necessary to extract the primary charged-particle multiplicity distributions are the raw, uncorrected measured multiplicity distributions and a response matrix $R$. The matrix $R$ is constructed via simulations where the known primary generated charged-particle multiplicity $T$ is correlated with the simulated detector response $M^{s}$. Figure~\ref{response-matrix} shows a graphical representation of response matrices obtained with the HIJING event generator for the two pseudorapidity coverages: $|\etalab|<2.4$ (left) and $-3.4<\etalab<5.0$ (right). The simulated detector response takes into account known conditions at the time of the data-taking, including inefficiencies, acceptance, electronic noise, and other smearing effects. Thus, one can write $M^{s} \approx RT$. The matrix element $R_{mt}$ represents the conditional probability that an event with true multiplicity $t$ is measured as an event with multiplicity $m$.

Experimentally, one needs to determine $T$ for a given measured charged-particle distribution $M$. This can be symbolically
written as
\begin{equation}
  T=R^{-1}M.
  \label{matrix_inversion}
\end{equation}

However, the matrix $R$ may be singular and cannot always be inverted analytically. Furthermore, even if $R$ can be inverted, the results obtained with Eq.~\eqref{matrix_inversion} contain oscillations mainly because of finite statistics in the response matrix. A regularised unfolding method based on Bayes' theorem~\cite{BayesUnfolding} using the RooUnfold software package~\cite{roounfold} is used to overcome this problem.

The Bayesian unfolding technique is an iterative method in which the number of iterations serves as a regularisation parameter. Given an initial hypothesis (a prior), $P_t$, with $t$ = 1, ..., $n$, for the true distributions, Bayes' theorem provides an estimation of the inverse matrix elements, $\tilde{R}_{tm}$,

\begin{equation*}
  \tilde{R}_{tm} = \frac{R_{mt}P_t}{\sum_{t^{'}} R_{mt^{'}}P_{t^{'}}}.
  \label{bayestheorem}
\end{equation*}

The unfolded distribution, $U_t$ , is then obtained from

\begin{equation*}
  U_t = \sum_{m} \tilde{R}_{tm} M_m.
  \label{unfoldeqn}
\end{equation*}

The obtained $U_t$ is used as the prior distribution for the next iteration. After each iteration, the iterative process makes the unfolded distribution closer to the true one. In order to optimise the number of iterations, the $\chi^2/$ndf between the unfolded and the true distribution is computed and then studied as a function of the number of iterations using MC simulations. The number of iterations is then set to the number for which the $\chi^2/$ndf becomes minimum. The optimised number of iterations is found to be from 2 to 3 for the different pseudorapidity ranges. These number of iterations are used to unfold the experimental data to obtain the corrected multiplicity distributions.

The unfolded distributions are corrected further for the collision selection efficiency ($\epsilon$), estimated via simulations as:

\begin{equation*}
\epsilon = \frac{N_{\rm detected}}{N_{\rm simulated}},
\end{equation*}
where $N_{\rm detected}$ is the number of collisions detected by the simulated detector using NSD trigger condition with $|z_{\mathrm{vtx}}|<4$\,cm and $N_{\rm simulated}$ is the number of simulated NSD collisions with $|z_{\mathrm{vtx}}|<4$\,cm. There is a dependence in the $z_{\mathrm{vtx}}$ distribution and selecting $z_{\mathrm{vtx}}$ introduces a bias in the efficiency. The effect is visible only for narrow vertex selections, and it is not relevant for $|z_{\mathrm{vtx}}|<4$\,cm~\cite{ALICE:ppFrdChMultpaper}. The values of $\epsilon$ are estimated as a function of the primary charged-particle multiplicity ($N_{\rm ch}$). For the widest pseudorapidity range ($-3.4<\etalab<5.0$), the efficiency, $\epsilon$, is found to vary from 0.2 ($N_{\rm ch} \simeq 1$) to 0.9 ($N_{\rm ch} \simeq 15$) while for $|\etalab|<2.4$, $\epsilon$ varies from 0.6 ($N_{\rm ch} \simeq 1$) to 0.9 ($N_{\rm ch} \simeq 15$). For all the studied pseudorapidity intervals, $\epsilon$ tends to be 1 above $N_{\rm ch}>20$. The unfolded results are corrected by dividing the content of each multiplicity bin by its $\epsilon$ value.

\begin{table}[h!]
  \small
  \begin{center}
    \caption{Contributions to systematic uncertainties (in percent) in the measurements of multiplicity distributions of primary charged particles for different pseudorapidity intervals in \pPb collisions at \fivenn. Numbers are given at three characteristic multiplicity values of 2, the mean $\langle N_{\rm ch} \rangle$, and the value for which P($N_{\rm ch}$) = 10$^{-3}$. Where the uncertainty is less than 0.1\%, it is specified as `negl.' in the table.}
    \label{table:sysunc1}
    \renewcommand{\arraystretch}{1.2}
    \begin{tabular}{|c|c c c|c c c|c c c|c c c|}
      \hline
      \multirow{5}{*}{Sources} & \multicolumn{3}{c|}{$-3.4<\etalab<5.0$} & \multicolumn{3}{c|}{$-3.4<\etalab<3.4$} & \multicolumn{3}{c|}{$-3.0<\etalab<3.0$} & \multicolumn{3}{c|}{$-2.4<\etalab<2.4$}\\
      \cline{2-13}
      & \begin{turn}{90} $N_{\rm ch}=2$ \end{turn} & \begin{turn}{90} $N_{\rm ch}=\langle N_{\rm ch}\rangle$ \end{turn} & \begin{turn}{90} P($N_{\rm ch}$) = 10$^{\text-3}$ \end{turn} & \begin{turn}{90} $N_{\rm ch}=2$ \end{turn} & \begin{turn}{90} $N_{\rm ch}=\langle N_{\rm ch}\rangle$ \end{turn} & \begin{turn}{90} P($N_{\rm ch}$) = 10$^{\text-3}$ \end{turn} & \begin{turn}{90} $N_{\rm ch}=2$ \end{turn} & \begin{turn}{90} $N_{\rm ch}=\langle N_{\rm ch}\rangle$ \end{turn} & \begin{turn}{90} P($N_{\rm ch}$) = 10$^{\text-3}$ \end{turn} & \begin{turn}{90} $N_{\rm ch}=2$ \end{turn} & \begin{turn}{90} $N_{\rm ch}=\langle N_{\rm ch}\rangle$ \end{turn} & \begin{turn}{90} P($N_{\rm ch}$) = 10$^{\text-3}$ \end{turn} \\
      \hline
      Upstream & \multirow{2}{*}{9.7} & \multirow{2}{*}{0.8} & \multirow{2}{*}{3.7} & \multirow{2}{*}{9.8} & \multirow{2}{*}{0.7} & \multirow{2}{*}{4.3} & \multirow{2}{*}{9.7} & \multirow{2}{*}{0.7} & \multirow{2}{*}{4.4} & \multirow{2}{*}{2.6} & \multirow{2}{*}{0.3} & \multirow{2}{*}{2.8} \\
      material & & & & & & & & & & & & \\
      \hline
      Event & \multirow{3}{*}{20.0} & \multirow{3}{*}{0.3} & \multirow{3}{*}{0.5} & \multirow{3}{*}{21.6} & \multirow{3}{*}{0.2} & \multirow{3}{*}{0.4} & \multirow{3}{*}{18.9} & \multirow{3}{*}{0.2} & \multirow{3}{*}{0.5} & \multirow{3}{*}{1.6} & \multirow{3}{*}{0.2} & \multirow{3}{*}{0.5} \\
      generator & & & & & & & & & & & & \\
      dependence & & & & & & & & & & & & \\
      \hline
      Unfolding & \multirow{2}{*}{6.5} & \multirow{2}{*}{0.1} & \multirow{2}{*}{0.1} & \multirow{2}{*}{4.4} & \multirow{2}{*}{negl.} & \multirow{2}{*}{negl.} & \multirow{2}{*}{1.8} & \multirow{2}{*}{negl.} & \multirow{2}{*}{negl.} & \multirow{2}{*}{4.5} & \multirow{2}{*}{negl.} & \multirow{2}{*}{0.1} \\
      parameters & & & & & & & & & & & & \\
      \hline
      Collision & \multirow{3}{*}{29.3} & \multirow{3}{*}{negl.} & \multirow{3}{*}{negl.} & \multirow{3}{*}{23.3} & \multirow{3}{*}{negl.} & \multirow{3}{*}{negl.} & \multirow{3}{*}{17.2} & \multirow{3}{*}{negl.} & \multirow{3}{*}{negl.} & \multirow{3}{*}{8.1} & \multirow{3}{*}{negl.} & \multirow{3}{*}{negl.} \\
      selection & & & & & & & & & & & & \\
      efficiency & & & & & & & & & & & & \\
      \hline
      Charged-particle & \multirow{3}{*}{4.0} & \multirow{3}{*}{0.3} & \multirow{3}{*}{3.1} & \multirow{3}{*}{3.4} & \multirow{3}{*}{0.3} & \multirow{3}{*}{2.7} & \multirow{3}{*}{2.6} & \multirow{3}{*}{0.2} & \multirow{3}{*}{2.5} & \multirow{3}{*}{1.2} & \multirow{3}{*}{0.2} & \multirow{3}{*}{1.8} \\ 
      detection & & & & & & & & & & & & \\
      thresholds & & & & & & & & & & & & \\
      \hline
      Total & 37.6 & 0.9 & 5.0 & 34.0 & 0.8 & 5.1 & 27.5 & 0.8 & 5.1 & 9.8 & 0.4 & 3.4 \\
      \hline
    \end{tabular}
  \end{center}
\end{table}

\begin{table}[h!]
  \small
  \begin{center}
    \caption{Contributions to systematic uncertainties (in percent) in the measurements of multiplicity distributions of primary charged particles on the p-fragmentation and the Pb-fragmentation sides in \pPb collisions at \fivenn. Numbers are given at three characteristic multiplicity values of 2, the mean $\langle N_{\rm ch} \rangle$, and the value for which P($N_{\rm ch}$) = 10$^{-3}$. Where the uncertainty is less than 0.1\%, it is specified as `negl.' in the table.}
    \label{table:sysunc2}
    \renewcommand{\arraystretch}{1.2}
    \begin{tabular}{|c|c c c|c c c|}
      \hline
      \multirow{5}{*}{Sources} & \multicolumn{3}{c|}{$-3.4<\etalab<-1.0$} & \multicolumn{3}{c|}{$2.0<\etalab<5.0$} \\
      & \multicolumn{3}{c|}{(p-fragmentation side)} & \multicolumn{3}{c|}{(Pb-fragmentation side)} \\ 
      \cline{2-7}
      & \begin{turn}{90} $N_{\rm ch}=2$ \end{turn} & \begin{turn}{90} $N_{\rm ch}=\langle N_{\rm ch}\rangle$ \end{turn} & \begin{turn}{90} P($N_{\rm ch}$) = 10$^{\text-3}$ \end{turn} & \begin{turn}{90} $N_{\rm ch}=2$ \end{turn} & \begin{turn}{90} $N_{\rm ch}=\langle N_{\rm ch}\rangle$ \end{turn} & \begin{turn}{90} P($N_{\rm ch}$) = 10$^{\text-3}$ \end{turn} \\
      \hline
      Upstream & \multirow{2}{*}{1.5} & \multirow{2}{*}{1.1} & \multirow{2}{*}{7.5} & \multirow{2}{*}{3.4} & \multirow{2}{*}{1.0} & \multirow{2}{*}{9.2} \\
      material & & & & & & \\
      \hline
      Event & \multirow{3}{*}{4.9} & \multirow{3}{*}{0.4} & \multirow{3}{*}{1.3} & \multirow{3}{*}{4.2} & \multirow{3}{*}{0.4} & \multirow{3}{*}{3.5} \\
      generator & & & & & & \\
      dependence & & & & & & \\
      \hline
      Unfolding & \multirow{2}{*}{1.7} & \multirow{2}{*}{negl.} & \multirow{2}{*}{0.2} & \multirow{2}{*}{2.7} & \multirow{2}{*}{0.1} & \multirow{2}{*}{0.1} \\
      parameters & & & & & & \\
      \hline
      Collision & \multirow{3}{*}{1.0} & \multirow{3}{*}{negl.} & \multirow{3}{*}{negl.} & \multirow{3}{*}{8.9} & \multirow{3}{*}{negl.} & \multirow{3}{*}{negl.} \\
      selection & & & & & & \\
      efficiency & & & & & & \\
      \hline
      Charged-particle & \multirow{3}{*}{1.7} & \multirow{3}{*}{0.3} & \multirow{3}{*}{1.4} & \multirow{3}{*}{2.2} & \multirow{3}{*}{1.2} & \multirow{3}{*}{3.7} \\ 
      detection & & & & & & \\
      thresholds & & & & & & \\
      \hline
      Total & 5.7 & 1.0 & 7.7 & 11.0 & 1.6 & 10.5 \\
      \hline
    \end{tabular}
  \end{center}
\end{table}

\section{Systematic uncertainties}
\label{SysUncEst}
The different sources of systematic uncertainties associated with the present measurements are summarised in Tables~\ref{table:sysunc1} and~\ref{table:sysunc2}. The first four contributions (upstream material, event generator dependence, unfolding parameters, collision selection efficiency) are common systematic uncertainties shared by both the SPD and the FMD while the last one (charged-particle detection thresholds) is only related to the FMD. The uncertainties vary with multiplicity; therefore, they are reported for three characteristic multiplicity values: $N_{\rm ch}$ = 2, the mean $\langle N_{\rm ch}\rangle$, and the value for which P($N_{\rm ch}$) = 10$^{-3}$, i.e. in the low, middle and high range, respectively. The total systematic uncertainty is calculated as the square root of the quadratic sum of the individual uncertainties (briefly described below).

The first source of systematic uncertainty arises from the uncertainty in the description of upstream material, between the nominal IP and the SPD and FMD, in the experimental simulations. The material in front of the detectors is a source of secondary particles which must be corrected for to obtain the primary particle distributions. The possibility to form tracklets from the SPD measurements is an effective way of disentangling the primary particle signal from the background from secondary particles. Therefore, only a small residual correction, with associated systematic uncertainty is needed at midrapidity. At forward rapidities, there is no possibility to form tracklets. This fact coupled with a large amount of material in front of the FMD makes it crucial to accurately simulate the production of secondary particles. However, there is considerable uncertainty in the description of the material in the detector simulations. Therefore, two sets of simulations are performed: one where all material densities are decreased by 5\% and another where these are increased by 10\%. Along with the nominal simulations, these two simulations probe the unknown distribution of possible secondary particle production in the material. We then apply a rigorous method~\cite{Barlow:2003sg} (where the asymmetric variations in material densities are treated as the standard deviations of two halved Gaussian functions and the resulting uncertainty is obtained by constructing a distorted Gaussian with the corresponding mean, variance, and skewness derived from those two halved Gaussians) to estimate the variance of that unknown distribution and assign that as a systematic uncertainty due to the imprecise knowledge of the material in front of the detectors.

To determine the systematic uncertainty due to the event generator's dependence on the unfolding procedure, the measured distributions in data are unfolded using two separate response matrices built using HIJING and DPMJET. The average of these two unfolded distributions is used as our final measurement. The resulting difference between the average value and the unfolded distributions obtained using HIJING and DPMJET is assigned as the systematic uncertainty. As described in Sec.~\ref{correction}, the unfolding of measured distributions is sensitive to the choice of the number of iterations in the Bayesian unfolding procedure. To account for this, unfolded distributions are obtained by varying the number of iterations by $\pm$\,1 around the optimised values. The deviations of these modified unfolded results from the nominal ones are considered as the systematic uncertainty.

The systematic uncertainty associated with the correction for the collision selection efficiency is evaluated by determining the efficiency values using two different event generators, HIJING and DPMJET. This uncertainty is largest at low multiplicity values and reduces significantly at larger $N_{\rm ch}$ because contributions from diffractive processes become smaller when going to higher multiplicity~\cite{ALICE:ppMidChMultpaper900and2360,ALICE:ppMidChMultpaper900to8000,ALICE:ppFrdChMultpaper}.

Depending on the incident angle, a charged particle may deposit energy in more than one FMD strip~\cite{ALICE:PbPbsatellite}. Signals shared in the neighbouring strips are then merged based on specific thresholds: a lower threshold ($T_{\rm low}$) for accepting a signal and an upper threshold ($T_{\rm high}$) to consider a signal as isolated, i.e. all energy is deposited in a single strip. The lower threshold is defined by the noise level ($n$) in the detector as $T_{\rm low} = xn$, where the factor $x$ is typically varied by one unit to estimate the one sigma variance in $N_{\rm ch}$. The upper threshold is set such that the probability of energy loss ($\Delta$) exceeding $T_{\rm high}$ for a single minimum ionizing particle (1 MIP) is greater than 99\% ($P(\Delta > T_{\rm high}|1\rm{MIP}) > 99$\%). This threshold is varied so that the probability increases or decreases by one standard deviation, thus estimating the variance of $N_{\rm ch}$. In order to calculate the number of charged particles using a Poisson statistical approach, the strips in the FMD are divided into regions, and the number of empty strips is compared to the total number of strips in a given region. Strips with a signal below a given threshold are considered empty. This threshold is varied within the boundaries of fits to the energy loss spectra to evaluate the systematic uncertainty.

\section{Results and discussion}
\label{results}
\begin{figure}[h!]
  \centering
  \includegraphics[scale=0.7]{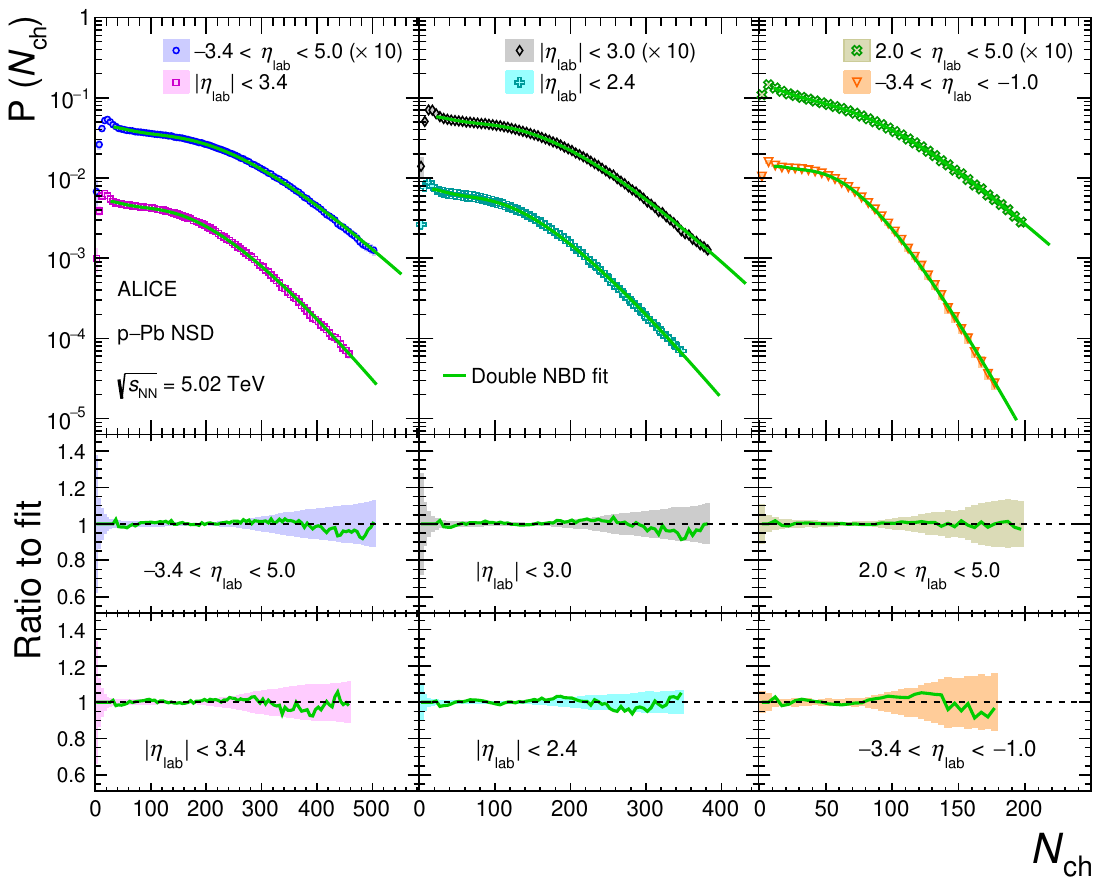}
  \caption{Charged-particle multiplicity distributions for different pseudorapidity intervals measured in \pPb collisions at \fivenn for NSD collisions. The green lines show fits of a double NBD function to the data. The ratios of the data to the fits are shown in the bottom panels.}
  \label{MultDistNBDfit}
\end{figure}

The primary charged-particle multiplicity distributions are measured for NSD \pPb collisions at \linebreak \snn = 5.02 TeV in six bins of pseudorapidities: $-3.4<\etalab<5.0$, $|\etalab|<3.4$, $|\etalab|<3.0$, $|\etalab|<2.4$, $-3.4<\etalab<-1.0$, and $2.0<\etalab<5.0$. The results are presented in Fig.~\ref{MultDistNBDfit}. In the widest pseudorapidity range, the multiplicity distribution reaches a maximum around $N_{\rm ch}\approx22$, while for $|\etalab|<2.4$, the maximum occurs around $N_{\rm ch}\approx12$. Beyond the maxima, the distributions fall steeply over several orders of magnitude. The coloured bands represent the systematic uncertainties, and the statistical uncertainties are smaller than the marker size. The multiplicity distributions P($N_{\rm ch}$) are found to broaden as the \etalab range increases. These measurements extend the high-multiplicity reach with respect to the previous ALICE results of \pp collisions both in the central~\cite{ALICE:ppMidChMultpaper900and2360,ALICE:ppMidChMultpaper7000,ALICE:ppMidChMultpaper900to8000} and forward rapidity~\cite{ALICE:ppFrdChMultpaper} regions. The distributions for $-3.4<\etalab<5.0$, $|\etalab|<3.0$, and $2.0<\etalab<5.0$ are scaled by a factor of 10 for clarity. The green lines show fits using a double Negative Binomial Distribution (NBD) function to the data, as discussed in the next subsection.

\subsection{Parametrisation of multiplicity distributions with double NBDs}
\label{doubleNBD}
Experimental measurements in \pp (\ppbar) collisions at \s $\leq$ 2.36 TeV for charged particles at midrapidity ($|\eta|<1.5$)~\cite{ALICE:ppMidChMultpaper900and2360,UA5:1988gup,Ghosh:2012xh} have shown that the multiplicity distributions can be described by a single NBD given by the probability density function (p.d.f.)

\begin{equation*}
  f_{\rm NBD}(n;\langle n \rangle,k) = \frac{\Gamma(n+k)}{\Gamma(k)\Gamma(n+1)}\frac{(\langle n \rangle/k)^{n}}{(1 + \langle n \rangle/k)^{n+k}}.
  \label{singleNBDEqn}
\end{equation*}

Here, $\langle n \rangle$ denotes the mean multiplicity and the parameter $k$ is related to the standard deviation ($\sigma$) of the distribution by $\sigma/\langle n \rangle = \sqrt{1/\langle n \rangle + 1/k}$. However, at higher collision energies (\s $\geq$ 2.76 TeV) and wider pseudorapidity intervals ($-3.4<\eta<5.0$), such a description is not adequate~\cite{ALICE:ppMidChMultpaper7000,ALICE:ppMidChMultpaper900to8000,ALICE:ppFrdChMultpaper,Ghosh:2012xh}. Instead, those measurements are better captured by a double NBD p.d.f.~\cite{ALICE:ppMidChMultpaper900to8000,ALICE:ppFrdChMultpaper,Ghosh:2012xh} given by 
\begin{equation}
  g(n;\langle n \rangle_1,k_1,\langle n \rangle_2,k_2,\lambda,\alpha) = \lambda[\alpha f_{\rm NBD} (n;\langle n \rangle_1,k_1) +
    (1 - \alpha)f_{\rm NBD} (n;\langle n \rangle_2,k_2)].
  \label{DoubleNBDEqn}
\end{equation}

In Eq.~\eqref{DoubleNBDEqn}, $\langle n \rangle_1$ and $\langle n \rangle_2$ are the mean multiplicities of the first and second components (often interpreted as corresponding to soft and semihard processes), respectively, and the parameter $\alpha$ reflects the fraction of the first component~\cite{Giovannini:1998zb,Giovannini:1999tw,Ghosh:2012xh}. The parameters $k_1$ and $k_2$ are related to the standard deviations of the distributions associated with the first and second components, respectively.

In this work, the double NBD p.d.f. (given by Eq.~\eqref{DoubleNBDEqn}) is fitted to the measured multiplicity distributions. The first few bins ($N_{\rm ch}\approx$ 10 to 30 depending on $\eta$ window) of the multiplicity distributions are excluded from the fit and a free normalisation factor $\lambda$ is introduced to account for this. This cut-off is chosen, primarily, to avoid the low-multiplicity shape that is known to be not compatible with double NBD p.d.f.~\cite{ALICE:ppMidChMultpaper900to8000,ALICE:ppFrdChMultpaper}. Coincidentally, diffractive contribution is only present at low multiplicities, however, there are no specific measurements available to estimate the importance of its effect on the shape of the multiplicity distribution. Nevertheless, it is expected that this contribution is well below the chosen cut-off as high-mass diffractive events that contribute to central multiplicity are quite rare. The fits are plotted together with the measured distributions in Fig.~\ref{MultDistNBDfit}. The double NBD function reasonably describes the data within the uncertainties.

\begin{figure}[h!]
  \centering
  \includegraphics[scale=0.5]{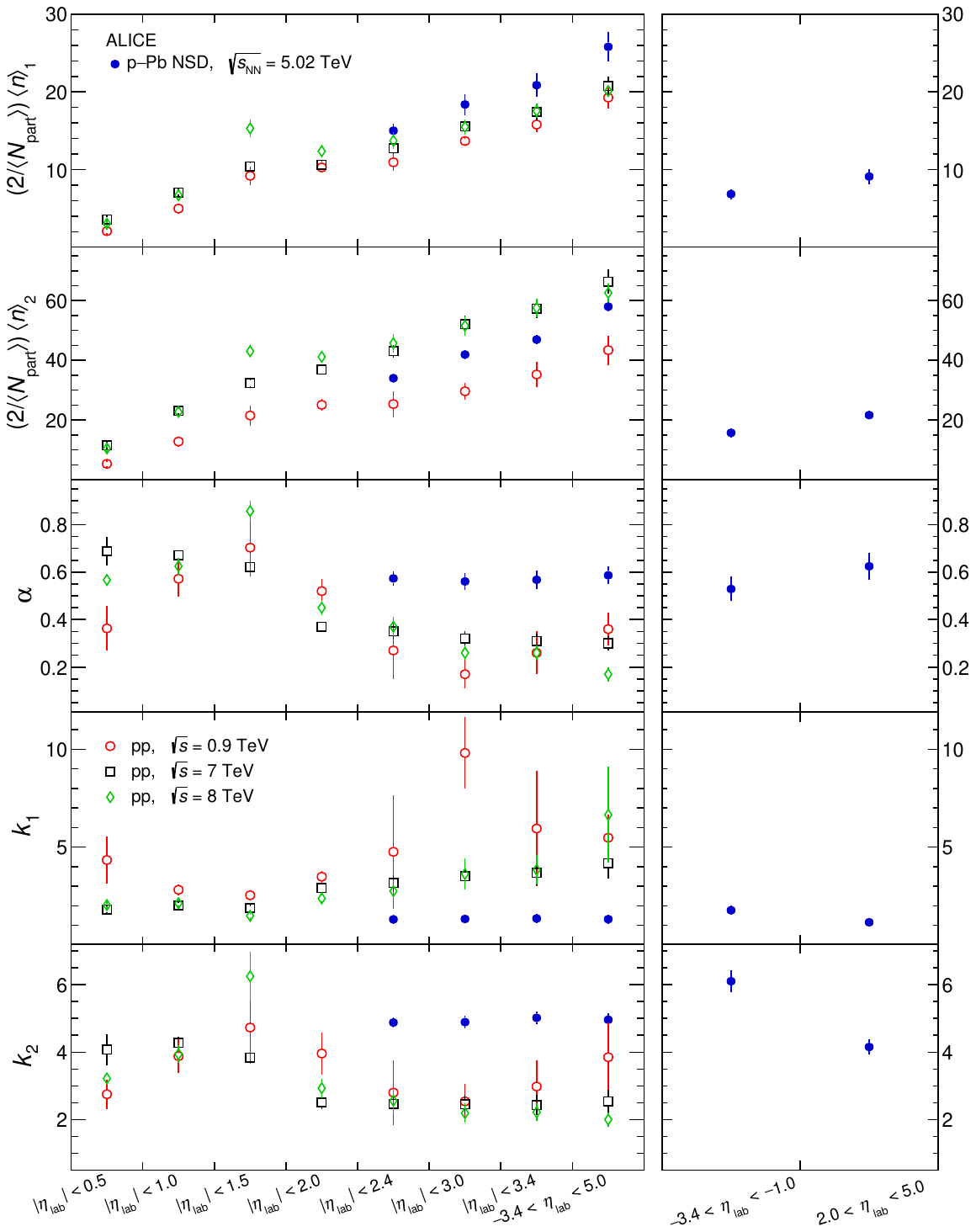}
  \caption{The pseudorapidity dependence of the double NBD parameters: $\langle n \rangle_1$, $\langle n \rangle_2$, $k_1$, and $k_2$ in \pPb collisions at \fivenn in comparison with \pp measurements at \s = 0.9, 7, and 8 \TeV~\cite{ALICE:ppMidChMultpaper900to8000,ALICE:ppFrdChMultpaper}. For \pPb collisions, the $\langle n \rangle_1$ and $\langle n \rangle_2$ values are scaled by the $\langle \Npart \rangle$/2.}
  \label{NBDPara_fig}
\end{figure}

The obtained parameters from the fit to the data for different pseudorapidity intervals are shown in Fig.~\ref{NBDPara_fig}. The fit parameters obtained in \pPb collisions are compared with the available \pp measurements~\cite{ALICE:ppMidChMultpaper900to8000,ALICE:ppFrdChMultpaper}. In \pPb collisions, values of $\langle n \rangle_1$ and $\langle n \rangle_2$ are normalised by the average number of participating nucleon pairs ($\langle \Npart \rangle$/2). Both $\langle n \rangle_1$ and $\langle n \rangle_2$ increase with the increase in \etalab. It is found that $\langle n \rangle_2 \simeq 3 \langle n \rangle_1$ for \pp collisions at \s = 7 and 8 TeV whereas $\langle n \rangle_2 \simeq 2.4 \langle n \rangle_1$ for \pp collisions at \s = 0.9 TeV and \pPb collisions at \fivenn. This observation suggests a relationship between the two components of the multiplicity distribution, which may reflect the relative contributions from soft and semihard processes. In the left panels of Fig.~\ref{NBDPara_fig}, one can notice that for increasing pseudorapidity ranges starting at $|\etalab|<2.4$, the normalised $\langle n \rangle_1^{\rm p-Pb} \gtrapprox \langle n \rangle_1^{\rm pp}$ whereas the normalised $\langle n \rangle_2^{\rm p-Pb}$ lies between the values observed at 0.9 and 7, 8 TeV for pp collisions. This suggests that the average multiplicity of the first (soft) component is nearly identical for both \pp and \pPb collisions, whereas the second (semihard) component follows an energy-dependent trend, increasing with energy. The parameters $\alpha$, $k_1$, and $k_2$ are independent of the width of the measured pseudorapidity interval for \pPb collisions unlike in \pp where they are found to have a mild dependence on the width of \etalab range. We observe clear differences in the NBD parameters of the multiplicity distributions between the p-fragmentation and the Pb-fragmentation sides (right panels of Fig.~\ref{NBDPara_fig}) in \pPb collisions. The Pb-fragmentation side exhibits higher values of $\langle n \rangle_1$ and $\langle n \rangle_2$, likely due to increased particle production relative to the p-fragmentation side. In addition, both $\langle k \rangle_1$ and $\langle k \rangle_2$ are found to decrease from the p-fragmentation to the Pb-fragmentation side, while the parameter $\alpha$ is approximately similar for both sides.

\subsection{Moments of the multiplicity distributions}
\label{MultDist}
\begin{figure}[h!]
  \centering
  \includegraphics[scale=0.5]{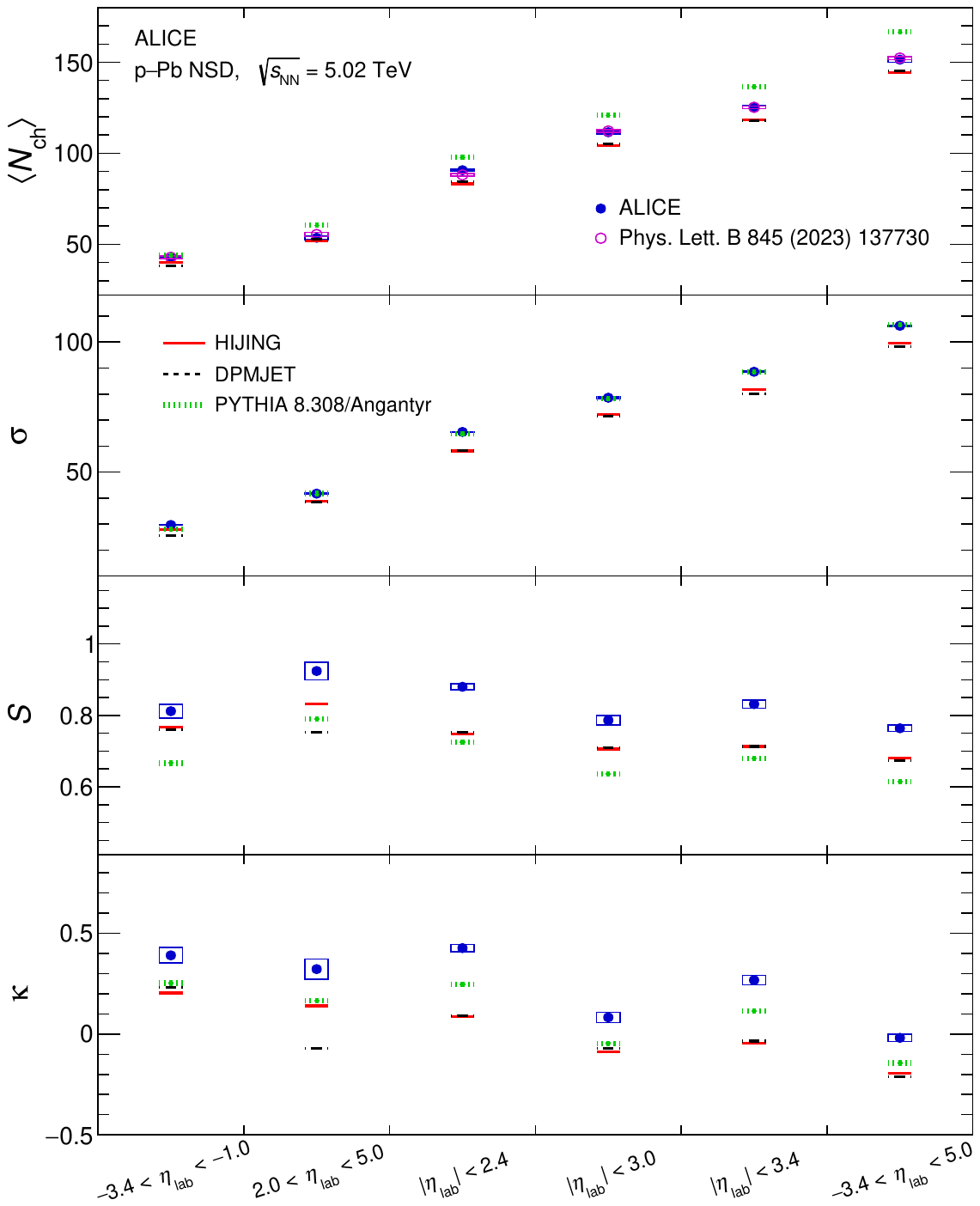}
  \caption{Four moments: $\langle N_{\rm ch}\rangle$, $\sigma$, $S$, and $\kappa$ of charged-particle multiplicity distributions for different pseudorapidity intervals in \pPb collisions at \fivenn. Both skewness and kurtosis are plotted on two different ordinate scales to better visualize their respective variations. Predictions from the HIJING, DPMJET, and \linebreak \angantyr event generators are superimposed.}
  \label{moments_fig}
\end{figure}

To study multiplicity distributions and their shape, the first four moments ($\langle N_{\rm ch}\rangle$, $\sigma$, $S$, and $\kappa$) are calculated. The obtained values of $\langle N_{\rm ch}\rangle$, $\sigma$, $S$, and $\kappa$ of the measured multiplicity distributions at different pseudorapidity intervals are shown in Fig.~\ref{moments_fig}. The open boxes represent the systematic uncertainty and the statistical errors are smaller than the symbols. The values of $\langle N_{\rm ch}\rangle$ and $\sigma$ rise with the increasing width of the pseudorapidity interval. The expectation values of $N_{\rm ch}$ are also compared to those derived from previous ALICE \dndeta measurements~\cite{ALICE:2022imr} (open circles), which differ in methodology, and, consequently, have different uncertainties, albeit with some overlap. Both measurements are found to be consistent and have uncertainties of less than 1\%, and overlapping uncertainties contribute no more than half of that uncertainty. The skewness is positive, showing only a modest variation of approximately 0.2 across the studied \etalab intervals, while the kurtosis exhibits a weakly decreasing trend with increasing \etalab interval, changing by about 0.5. The different lines in Fig.~\ref{moments_fig} are predictions from the HIJING, DPMJET, and \angantyr (default tune) event generators. The models follow the general trend of the data points; however, they show significant deviations from the data. The moments of the HIJING and DPMJET distributions are similar except the $S$ and $\kappa$ on the Pb-fragmentation side. The $\langle N_{\rm ch}\rangle$ of the HIJING and DPMJET distributions are close to the data, but for the higher moments, they describe the data poorly, implying that the shape of their distributions is different from the data. On the other hand, \angantyr reproduces the $\sigma$ of the measured distributions but cannot explain the rest of the moments (except the $\langle N_{\rm ch}\rangle$ of the data on the p-fragmentation side).

\begin{figure}[h!]
  \centering
  \includegraphics[scale=0.8]{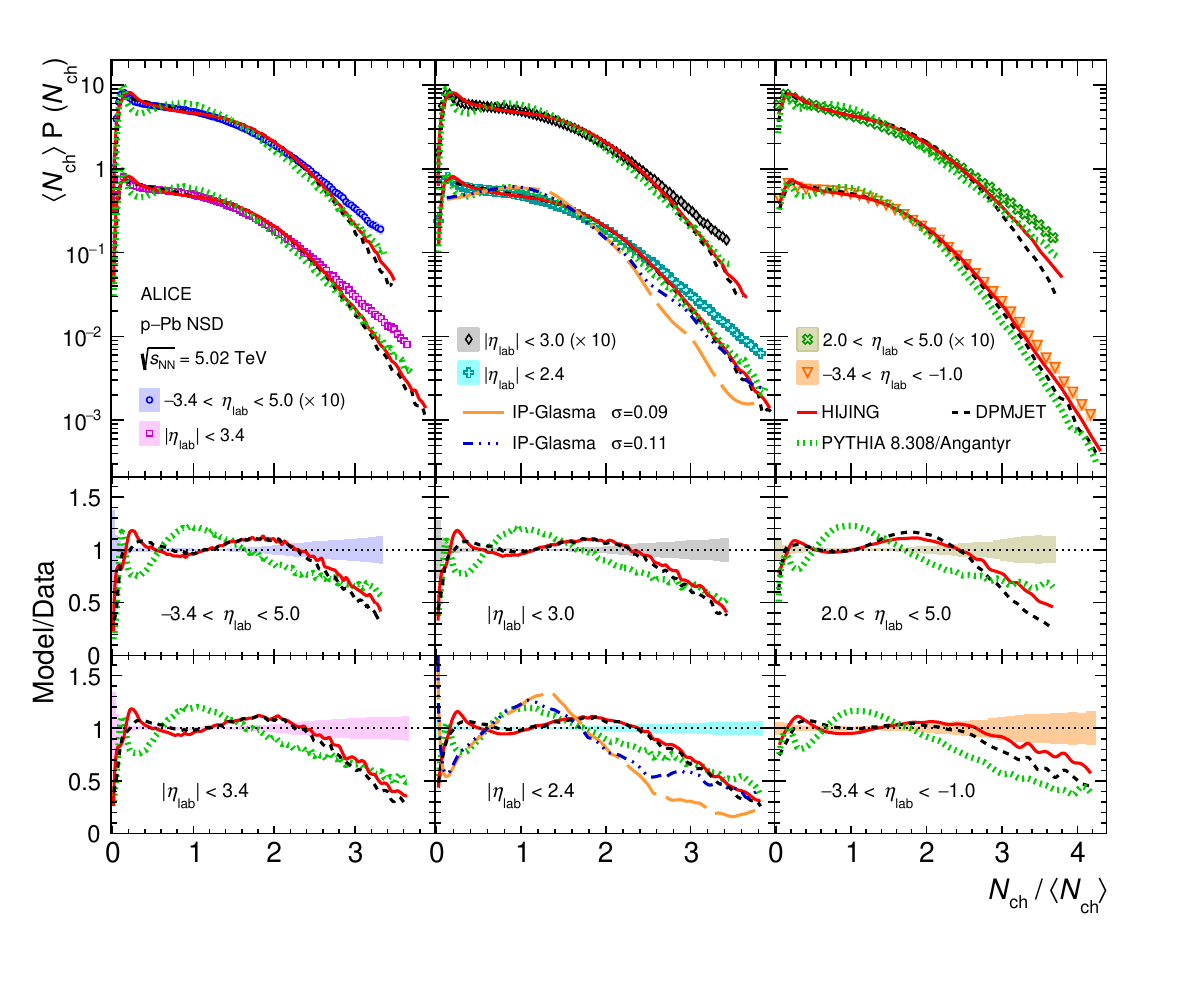}
  \caption{KNO-scaled multiplicity distribution versus the KNO variable $N_{\rm ch}/\langle N_{\rm ch}\rangle$ in NSD \pPb collisions at \fivenn for various pseudorapidity intervals. Comparison with predictions from HIJING, DPMJET, \angantyr, and the IP-Glasma model are shown. The ratios between models and data are calculated using a linear interpolation between adjacent points.}
  \label{MultDistMCcompare}
\end{figure}

\subsection{KNO scaling in the multiplicity distributions}
Koba, Nielsen and Olesen (KNO) found that for lower energy collisions, all moments of the multiplicity distribution scale with the first moment, i.e., $\langle N_{\rm ch}^{\rm n}\rangle \propto \langle N_{\rm ch}\rangle$~\cite{Koba:1972ng}. Thus, a way to investigate properties of the multiplicity distributions is to plot these scaled by the mean multiplicity using the so-called KNO variable $N_{\rm ch}/\langle N_{\rm ch}\rangle$. This also has the added benefit that models that may differ in the mean of the distribution can still be compared to the empirical data. Figure~\ref{MultDistMCcompare} presents the data after scaling the probability density and the charged-particle multiplicity with the average number of charged particles $\langle N_{\rm ch}\rangle$. The distributions for $-3.4<\etalab<5.0$, $|\etalab|<3.0$, and $2.0<\etalab<5.0$ are scaled by a factor of 10 for clarity. The data are compared with predictions from the HIJING, DPMJET, and \angantyr event generators. The models underestimate the data both at low and high multiplicities, indicating that they give narrower distributions than the data. The HIJING and DPMJET distributions are close to one another and compatible with the data (within 10\%) for $0.2<N_{\rm ch}/\langle N_{\rm ch}\rangle<2.5$. This indicates that HIJING and DPMJET provide similar $\langle N_{\rm ch}\rangle$ values relative to data (also evident in the top panel of Fig.~\ref{moments_fig}). On the other hand, \angantyr gives the poorest description of the data in the intermediate multiplicities than the other two MC models. More specifically, \angantyr is lower than the data for $0.2<N_{\rm ch}/\langle N_{\rm ch}\rangle<0.6$ while higher than the data for $0.6<N_{\rm ch}/\langle N_{\rm ch}\rangle<1.7$.

The measurement in $|\etalab|<2.4$ is also compared to the prediction from the IP-Glasma model~\cite{Schenke:2013dpa} based on the Color Glass Condensate (CGC) framework~\cite{CGCmodel}. The IP-Glasma model incorporates fluctuations in the density of colour charges. In Fig.~\ref{MultDistMCcompare}, the orange and blue distributions are generated with fluctuations of the colour charge density around the mean following a Gaussian distribution with width \linebreak $\sigma$ = 0.09 and 0.11, respectively. The IP-Glasma model, irrespective of the size of the fluctuations, largely overestimates the data at very low multiplicities ($N_{\rm ch}/\langle N_{\rm ch}\rangle<0.1$) and underestimates the same at high multiplicities ($N_{\rm ch}/\langle N_{\rm ch}\rangle>2$).

\subsection{System-size and energy dependence of \texorpdfstring{$\langle N_{\rm ch} \rangle$}{Nch}}
\begin{figure}[h!]
  \centering
  \includegraphics[scale=0.5]{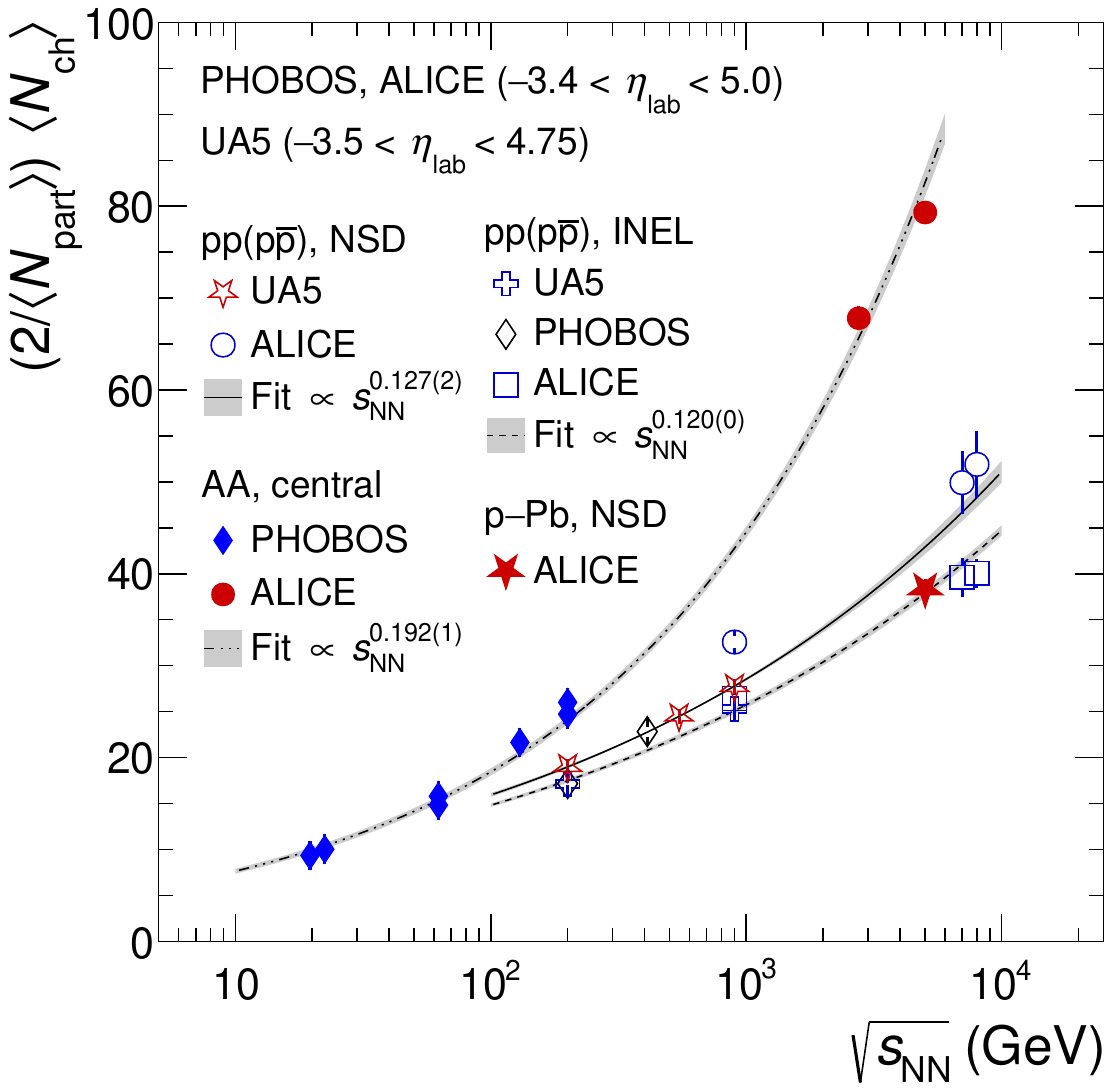}
  \caption{Values of $\frac{2}{\langle N_{\rm part} \rangle} \langle N_{\rm ch} \rangle$ for minimum-bias \pp~\cite{ALICE:ppFrdChMultpaper,PHOBOS:2010eyu}, \ppbar~\cite{UA5:1986ye}, \pPb and central AA~\cite{ALICE:PbPb_CentFrdrapidity2760,ALICE:PbPbsatellite,ALICE:PbPb_CentFrdrapidity5020} collisions as a function of \snn are shown. The $s_{\mathrm{NN}}$-dependencies of INEL \pp (\ppbar) and NSD \pp (\ppbar) collisions are proportional to $s_{\mathrm{NN}}^{0.120}$ and $s_{\mathrm{NN}}^{0.127}$ respectively. The results from central AA collisions are proportional to $s_{\mathrm{NN}}^{0.192}$. The bands represent the uncertainties on the extracted power-law dependencies.}
  \label{AvgMultRootS}
\end{figure}

In order to understand and compare the evolution of bulk particle production with collision energy and system-size, the mean charged-particle multiplicity is normalised by the $\langle \Npart \rangle$ pairs and then presented as a function of \snn in Fig.~\ref{AvgMultRootS} for different collision systems. The $\langle N_{\rm ch} \rangle$ is measured over a range more than eight units in pseudorapidity and the $\langle \Npart \rangle$ is estimated using Glauber model calculations~\cite{PHOBOS:2010eyu,ALICE:2013hur,ALICE:PbPb_CentFrdrapidity5020,ALICE:pPbMidChdNdEtapaper5020Cent}. Data from inelastic (INEL) and non-single diffractive \pp (\ppbar) collisions~\cite{ALICE:ppFrdChMultpaper,PHOBOS:2010eyu,UA5:1986ye} and central heavy-ion collisions~\cite{ALICE:PbPb_CentFrdrapidity2760,ALICE:PbPbsatellite,ALICE:PbPb_CentFrdrapidity5020} are shown for comparison. A power-law ($\alpha \cdot s_{\rm NN}^{\beta}$) is fitted to the $\langle N_{\rm ch} \rangle$ as a function of centre-of-mass energy. Best-fit parameter values are $\beta$ = 0.120 $\pm$ 0.0001, 0.127 $\pm$ 0.002, and 0.192 $\pm$ 0.001 for INEL \pp (\ppbar), NSD \pp (\ppbar), and central AA collisions, respectively. The fit results are presented with their uncertainties shown by shaded bands. The results clearly show that the normalised $\langle N_{\rm ch} \rangle$ increases faster with energy in central AA collisions than in \pp collisions. The value of $\frac{2}{\langle N_{\rm part} \rangle} \langle N_{\rm ch} \rangle$ measured in \pPb collisions at \fivenn is half the magnitude of that in \PbPb collisions at the same energy, and falls on the INEL \pp curve. A similar observation was also reported for charged-particle multiplicity measurements at midrapidity ($|\etalab|<0.5$)~\cite{ALICE:pPbMidChdNdEtapaper5020MB,ALICE:pPbMidChdNdEtapaper8160,CMS:ChPrMidpPb5TeV}. The similarity between the NSD \pPb and the INEL pp data is yet to be understood. 

\section{Summary}
\label{summary}
The multiplicity distributions of primary charged particles have been measured in non-single diffractive \pPb collisions at \fivenn using the ALICE detector at the LHC. The measurements were performed over a wide pseudorapidity range ($-3.4<\etalab<5.0$), the widest possible among the four large LHC experiments. The multiplicity distributions are parametrised with a double Negative Binomial Distribution function, which describes the data well within the measurement uncertainties. The first four moments (mean, standard deviation, skewness, and kurtosis) of the multiplicity distributions are determined and compared with predictions from the HIJING, DPMJET, and \angantyr MC event generators. HIJING and DPMJET describe the mean of the distribution within $\sim5\%$ but cannot explain the higher moments of the data. On the other hand, \angantyr reproduces only the second moment of the measured distributions but cannot describe the rest of the moments.

The multiplicity distributions are also presented as a function of the KNO variable and compared with predictions from HIJING, DPMJET, \angantyr, and the CGC-based IP-Glasma model. None of the models can reproduce the data in the reported multiplicity range. HIJING and DPMJET explain the data better than \angantyr in the intermediate multiplicities. However, all MC predictions largely underestimate the multiplicity distributions at low and high multiplicities. The CGC-based IP-Glasma model disagrees with the measurements, irrespective of the level of colour charge fluctuations introduced into that model.

Finally, the dependence of $\frac{2}{\langle N_{\rm part} \rangle} \langle N_{\rm ch} \rangle$ on the centre-of-mass energy is parametrised by a power-law function, which shows that the multiplicity in \pPb collisions coincides with the trend observed in inelastic pp collisions.

The measurements reported in this paper provide valuable information for better understanding particle production mechanisms in \pPb collisions and offer valuable input for developing theoretical models and Monte Carlo event generators.


\newenvironment{acknowledgement}{\relax}{\relax}
\begin{acknowledgement}
\section*{Acknowledgements}

The ALICE Collaboration would like to thank all its engineers and technicians for their invaluable contributions to the construction of the experiment and the CERN accelerator teams for the outstanding performance of the LHC complex.
The ALICE Collaboration gratefully acknowledges the resources and support provided by all Grid centres and the Worldwide LHC Computing Grid (WLCG) collaboration.
The ALICE Collaboration acknowledges the following funding agencies for their support in building and running the ALICE detector:
A. I. Alikhanyan National Science Laboratory (Yerevan Physics Institute) Foundation (ANSL), State Committee of Science and World Federation of Scientists (WFS), Armenia;
Austrian Academy of Sciences, Austrian Science Fund (FWF): [M 2467-N36] and Nationalstiftung f\"{u}r Forschung, Technologie und Entwicklung, Austria;
Ministry of Communications and High Technologies, National Nuclear Research Center, Azerbaijan;
Conselho Nacional de Desenvolvimento Cient\'{\i}fico e Tecnol\'{o}gico (CNPq), Financiadora de Estudos e Projetos (Finep), Funda\c{c}\~{a}o de Amparo \`{a} Pesquisa do Estado de S\~{a}o Paulo (FAPESP) and Universidade Federal do Rio Grande do Sul (UFRGS), Brazil;
Bulgarian Ministry of Education and Science, within the National Roadmap for Research Infrastructures 2020-2027 (object CERN), Bulgaria;
Ministry of Education of China (MOEC) , Ministry of Science \& Technology of China (MSTC) and National Natural Science Foundation of China (NSFC), China;
Ministry of Science and Education and Croatian Science Foundation, Croatia;
Centro de Aplicaciones Tecnol\'{o}gicas y Desarrollo Nuclear (CEADEN), Cubaenerg\'{\i}a, Cuba;
Ministry of Education, Youth and Sports of the Czech Republic, Czech Republic;
The Danish Council for Independent Research | Natural Sciences, the VILLUM FONDEN and Danish National Research Foundation (DNRF), Denmark;
Helsinki Institute of Physics (HIP), Finland;
Commissariat \`{a} l'Energie Atomique (CEA) and Institut National de Physique Nucl\'{e}aire et de Physique des Particules (IN2P3) and Centre National de la Recherche Scientifique (CNRS), France;
Bundesministerium f\"{u}r Bildung und Forschung (BMBF) and GSI Helmholtzzentrum f\"{u}r Schwerionenforschung GmbH, Germany;
General Secretariat for Research and Technology, Ministry of Education, Research and Religions, Greece;
National Research, Development and Innovation Office, Hungary;
Department of Atomic Energy Government of India (DAE), Department of Science and Technology, Government of India (DST), University Grants Commission, Government of India (UGC) and Council of Scientific and Industrial Research (CSIR), India;
National Research and Innovation Agency - BRIN, Indonesia;
Istituto Nazionale di Fisica Nucleare (INFN), Italy;
Japanese Ministry of Education, Culture, Sports, Science and Technology (MEXT) and Japan Society for the Promotion of Science (JSPS) KAKENHI, Japan;
Consejo Nacional de Ciencia (CONACYT) y Tecnolog\'{i}a, through Fondo de Cooperaci\'{o}n Internacional en Ciencia y Tecnolog\'{i}a (FONCICYT) and Direcci\'{o}n General de Asuntos del Personal Academico (DGAPA), Mexico;
Nederlandse Organisatie voor Wetenschappelijk Onderzoek (NWO), Netherlands;
The Research Council of Norway, Norway;
Pontificia Universidad Cat\'{o}lica del Per\'{u}, Peru;
Ministry of Science and Higher Education, National Science Centre and WUT ID-UB, Poland;
Korea Institute of Science and Technology Information and National Research Foundation of Korea (NRF), Republic of Korea;
Ministry of Education and Scientific Research, Institute of Atomic Physics, Ministry of Research and Innovation and Institute of Atomic Physics and Universitatea Nationala de Stiinta si Tehnologie Politehnica Bucuresti, Romania;
Ministry of Education, Science, Research and Sport of the Slovak Republic, Slovakia;
National Research Foundation of South Africa, South Africa;
Swedish Research Council (VR) and Knut \& Alice Wallenberg Foundation (KAW), Sweden;
European Organization for Nuclear Research, Switzerland;
Suranaree University of Technology (SUT), National Science and Technology Development Agency (NSTDA) and National Science, Research and Innovation Fund (NSRF via PMU-B B05F650021), Thailand;
Turkish Energy, Nuclear and Mineral Research Agency (TENMAK), Turkey;
National Academy of  Sciences of Ukraine, Ukraine;
Science and Technology Facilities Council (STFC), United Kingdom;
National Science Foundation of the United States of America (NSF) and United States Department of Energy, Office of Nuclear Physics (DOE NP), United States of America.
In addition, individual groups or members have received support from:
Czech Science Foundation (grant no. 23-07499S), Czech Republic;
FORTE project, reg.\ no.\ CZ.02.01.01/00/22\_008/0004632, Czech Republic, co-funded by the European Union, Czech Republic;
European Research Council (grant no. 950692), European Union;
Deutsche Forschungs Gemeinschaft (DFG, German Research Foundation) ``Neutrinos and Dark Matter in Astro- and Particle Physics'' (grant no. SFB 1258), Germany;
ICSC - National Research Center for High Performance Computing, Big Data and Quantum Computing and FAIR - Future Artificial Intelligence Research, funded by the NextGenerationEU program (Italy).
\end{acknowledgement}

\newpage
\bibliographystyle{utphys}   
\bibliography{bibliography}

\providecommand{\href}[2]{#2}\begingroup\raggedright\begin{thebibliography}{10}

\bibitem{ALICE:ppMidChMultpaper900and2360}
{\bfseries ALICE} Collaboration, K.~Aamodt {\em et~al.}, ``{Charged-particle
  multiplicity measurement in proton--proton collisions at $\sqrt{s}=0.9$ and
  2.36~TeV with ALICE at LHC}'',
  \href{https://doi.org/10.1140/epjc/s10052-010-1339-x}{{\em Eur. Phys. J. C}
  {\bfseries 68} (2010) 89--108},
  \href{https://arxiv.org/abs/1004.3034}{{\ttfamily arXiv:1004.3034 [hep-ex]}}.

\bibitem{ALICE:ppMidChMultpaper7000}
{\bfseries ALICE} Collaboration, K.~Aamodt {\em et~al.}, ``{Charged-particle
  multiplicity measurement in proton--proton collisions at
  $\sqrt{s}~=~7$~Te\kern-.1emV\xspace with ALICE at LHC}'',
  \href{https://doi.org/10.1140/epjc/s10052-010-1350-2}{{\em Eur. Phys. J. C}
  {\bfseries 68} (2010) 345--354},
  \href{https://arxiv.org/abs/1004.3514}{{\ttfamily arXiv:1004.3514 [hep-ex]}}.

\bibitem{ALICE:ppMidChMultpaper900to8000}
{\bfseries ALICE} Collaboration, J.~Adam {\em et~al.}, ``{Charged-particle
  multiplicities in proton\textendash{}proton collisions at $\sqrt{s} = 0.9$ to
  8~TeV}'', \href{https://doi.org/10.1140/epjc/s10052-016-4571-1}{{\em Eur.
  Phys. J. C} {\bfseries 77} (2017) 33},
  \href{https://arxiv.org/abs/1509.07541}{{\ttfamily arXiv:1509.07541
  [nucl-ex]}}.

\bibitem{ALICE:ppFrdChMultpaper}
{\bfseries ALICE} Collaboration, S.~Acharya {\em et~al.}, ``{Charged-particle
  multiplicity distributions over a wide pseudorapidity range in proton--proton
  collisions at $\sqrt{s}~=$~0.9, 7, and 8~TeV}'',
  \href{https://doi.org/10.1140/epjc/s10052-017-5412-6}{{\em Eur. Phys. J. C}
  {\bfseries 77} (2017) 852},
  \href{https://arxiv.org/abs/1708.01435}{{\ttfamily arXiv:1708.01435
  [hep-ex]}}.

\bibitem{ALICE:PbPbsatellite}
{\bfseries ALICE} Collaboration, E.~Abbas {\em et~al.}, ``{Centrality
  dependence of the pseudorapidity density distribution for charged particles
  in Pb--Pb collisions at
  $\sqrt{s_{\mathrm{NN}}}~=~2.76$~Te\kern-.1emV\xspace}'',
  \href{https://doi.org/10.1016/j.physletb.2013.09.022}{{\em Phys. Lett. B}
  {\bfseries 726} (2013) 610--622},
  \href{https://arxiv.org/abs/1304.0347}{{\ttfamily arXiv:1304.0347
  [nucl-ex]}}.

\bibitem{ALICE:PbPb_CentFrdrapidity2760}
{\bfseries ALICE} Collaboration, J.~Adam {\em et~al.}, ``{Centrality evolution
  of the charged-particle pseudorapidity density over a broad pseudorapidity
  range in Pb--Pb collisions at
  $\sqrt{s_{\mathrm{NN}}}~=~2.76$~Te\kern-.1emV\xspace}'',
  \href{https://doi.org/10.1016/j.physletb.2015.12.082}{{\em Phys. Lett. B}
  {\bfseries 754} (2016) 373--385},
  \href{https://arxiv.org/abs/1509.07299}{{\ttfamily arXiv:1509.07299
  [nucl-ex]}}.

\bibitem{ALICE:PbPb_CentFrdrapidity5020}
{\bfseries ALICE} Collaboration, J.~Adam {\em et~al.}, ``{Centrality dependence
  of the pseudorapidity density distribution for charged particles in Pb--Pb
  collisions at $\sqrt{s_{\mathrm{NN}}}~=~5.02$~Te\kern-.1emV\xspace}'',
  \href{https://doi.org/10.1016/j.physletb.2017.07.017}{{\em Phys. Lett. B}
  {\bfseries 772} (2017) 567--577},
  \href{https://arxiv.org/abs/1612.08966}{{\ttfamily arXiv:1612.08966
  [nucl-ex]}}.

\bibitem{ALICE:2022imr}
{\bfseries ALICE} Collaboration, S.~Acharya {\em et~al.}, ``{System-size
  dependence of the charged-particle pseudorapidity density at
  $\sqrt{{{s}}_{\textrm{NN}}}$ = 5.02 TeV for pp, p--Pb, and Pb--Pb
  collisions}'', \href{https://doi.org/10.1016/j.physletb.2023.137730}{{\em
  Phys. Lett. B} {\bfseries 845} (2023) 137730},
  \href{https://arxiv.org/abs/2204.10210}{{\ttfamily arXiv:2204.10210
  [nucl-ex]}}.

\bibitem{ALICE:RidgepPb}
{\bfseries ALICE} Collaboration, B.~Abelev {\em et~al.}, ``{Long-range angular
  correlations on the near and away side in p--Pb collisions at
  $\sqrt{s_{\mathrm{NN}}}$ = 5.02 TeV}'',
  \href{https://doi.org/10.1016/j.physletb.2013.01.012}{{\em Phys. Lett. B}
  {\bfseries 719} (2013) 29--41},
  \href{https://arxiv.org/abs/1212.2001}{{\ttfamily arXiv:1212.2001
  [nucl-ex]}}.

\bibitem{ATLAS:2012cix}
{\bfseries ATLAS} Collaboration, G.~Aad {\em et~al.}, ``{Observation of
  associated near-side and away-side long-range correlations in $\sqrt{s_{\rm
  NN}}$ = 5.02 TeV p--Pb Collisions with the ATLAS Detector}'',
  \href{https://doi.org/10.1103/PhysRevLett.110.182302}{{\em Phys. Rev. Lett.}
  {\bfseries 110} (2013) 182302},
  \href{https://arxiv.org/abs/1212.5198}{{\ttfamily arXiv:1212.5198 [hep-ex]}}.

\bibitem{CMS:2012qk}
{\bfseries CMS} Collaboration, S.~Chatrchyan {\em et~al.}, ``{Observation of
  long-range near-side angular correlations in p--Pb collisions at the LHC}'',
  \href{https://doi.org/10.1016/j.physletb.2012.11.025}{{\em Phys. Lett. B}
  {\bfseries 718} (2013) 795--814},
  \href{https://arxiv.org/abs/1210.5482}{{\ttfamily arXiv:1210.5482
  [nucl-ex]}}.

\bibitem{ATLAS:2014qaj}
{\bfseries ATLAS} Collaboration, G.~Aad {\em et~al.}, ``{Measurement of
  long-range pseudorapidity correlations and azimuthal harmonics in
  $\sqrt{s_{\rm NN}}=5.02$ TeV p--Pb collisions with the ATLAS detector}'',
  \href{https://doi.org/10.1103/PhysRevC.90.044906}{{\em Phys. Rev. C}
  {\bfseries 90} (2014) 044906},
  \href{https://arxiv.org/abs/1409.1792}{{\ttfamily arXiv:1409.1792 [hep-ex]}}.

\bibitem{ALICE:ReviewPaper}
{\bfseries ALICE} Collaboration, S.~Acharya {\em et~al.}, ``{The ALICE
  experiment: a journey through QCD}'',
  \href{https://doi.org/10.1140/epjc/s10052-024-12935-y}{{\em Eur. Phys. J. C}
  {\bfseries 84} (2024) 813},
  \href{https://arxiv.org/abs/2211.04384}{{\ttfamily arXiv:2211.04384
  [nucl-ex]}}.

\bibitem{Qiu:2004da}
J.-w. Qiu and I.~Vitev, ``{Coherent QCD multiple scattering in proton--nucleus
  collisions}'', \href{https://doi.org/10.1016/j.physletb.2005.10.073}{{\em
  Phys. Lett. B} {\bfseries 632} (2006) 507--511},
  \href{https://arxiv.org/abs/hep-ph/0405068}{{\ttfamily
  arXiv:hep-ph/0405068}}.

\bibitem{Wang:2001ifa}
X.-N. Wang and X.-f. Guo, ``{Multiple parton scattering in nuclei: parton
  energy loss}'', \href{https://doi.org/10.1016/S0375-9474(01)01130-7}{{\em
  Nucl. Phys. A} {\bfseries 696} (2001) 788--832},
  \href{https://arxiv.org/abs/hep-ph/0102230}{{\ttfamily
  arXiv:hep-ph/0102230}}.

\bibitem{hijing}
X.-N. Wang and M.~Gyulassy, ``{HIJING: A Monte Carlo model for multiple jet
  production in pp, p--A and AA collisions}'',
  \href{https://doi.org/10.1103/PhysRevD.44.3501}{{\em Phys. Rev. D} {\bfseries
  44} (1991) 3501--3516}.

\bibitem{dpmjet}
S.~Roesler, R.~Engel, and J.~Ranft,
  \href{https://doi.org/10.1007/978-3-642-18211-2_166}{``{The Monte Carlo event
  generator DPMJET-III}'',} in {\em {International Conference on Advanced Monte
  Carlo for Radiation Physics, Particle Transport Simulation and Applications
  (MC 2000)}}, pp.~1033--1038.
\newblock 12, 2000.
\newblock \href{https://arxiv.org/abs/hep-ph/0012252}{{\ttfamily
  arXiv:hep-ph/0012252 [hep-ph]}}.

\bibitem{Angantyr}
C.~Bierlich, G.~Gustafson, L.~L\"onnblad, and H.~Shah, ``{The Angantyr model
  for heavy-ion collisions in PYTHIA8}'',
  \href{https://doi.org/10.1007/JHEP10(2018)134}{{\em JHEP} {\bfseries 10}
  (2018) 134}, \href{https://arxiv.org/abs/1806.10820}{{\ttfamily
  arXiv:1806.10820 [hep-ph]}}.

\bibitem{Schenke:2012wb}
B.~Schenke, P.~Tribedy, and R.~Venugopalan, ``{Fluctuating glasma initial
  conditions and flow in heavy ion collisions}'',
  \href{https://doi.org/10.1103/PhysRevLett.108.252301}{{\em Phys. Rev. Lett.}
  {\bfseries 108} (2012) 252301},
  \href{https://arxiv.org/abs/1202.6646}{{\ttfamily arXiv:1202.6646
  [nucl-th]}}.

\bibitem{Schenke:2012hg}
B.~Schenke, P.~Tribedy, and R.~Venugopalan, ``{Event-by-event gluon
  multiplicity, energy density, and eccentricities in ultrarelativistic
  heavy-ion collisions}'',
  \href{https://doi.org/10.1103/PhysRevC.86.034908}{{\em Phys. Rev. C}
  {\bfseries 86} (2012) 034908},
  \href{https://arxiv.org/abs/1206.6805}{{\ttfamily arXiv:1206.6805 [hep-ph]}}.

\bibitem{ALICE:Exp}
{\bfseries ALICE} Collaboration, K.~Aamodt {\em et~al.}, ``{The ALICE
  experiment at the CERN LHC}'',
  \href{https://doi.org/10.1088/1748-0221/3/08/S08002}{{\em JINST} {\bfseries
  3} (2008) S08002}.

\bibitem{ALICE:performance}
{\bfseries ALICE} Collaboration, B.~B. Abelev {\em et~al.}, ``{Performance of
  the ALICE experiment at the CERN LHC}'',
  \href{https://doi.org/10.1142/S0217751X14300440}{{\em Int. J. Mod. Phys. A}
  {\bfseries 29} (2014) 1430044},
  \href{https://arxiv.org/abs/1402.4476}{{\ttfamily arXiv:1402.4476
  [nucl-ex]}}.

\bibitem{ALICE:FrdDettdr}
{\bfseries ALICE} Collaboration, P.~Cortese {\em et~al.}, ``{ALICE forward
  detectors: FMD, T0 and V0: Technical Design Report}'', CERN-LHCC-2004-025.
  \url{https://cds.cern.ch/record/781854}.

\bibitem{ALICE:V0performance}
{\bfseries ALICE} Collaboration, E.~Abbas {\em et~al.}, ``{Performance of the
  ALICE VZERO system}'',
  \href{https://doi.org/10.1088/1748-0221/8/10/P10016}{{\em JINST} {\bfseries
  8} (2013) P10016}, \href{https://arxiv.org/abs/1306.3130}{{\ttfamily
  arXiv:1306.3130 [nucl-ex]}}.

\bibitem{ALICE:ITStdr}
{\bfseries ALICE} Collaboration, K.~Aamodt {\em et~al.}, ``{Alignment of the
  ALICE Inner Tracking System with cosmic-ray tracks}'',
  \href{https://doi.org/10.1088/1748-0221/5/03/P03003}{{\em JINST} {\bfseries
  5} (2010) P03003}, \href{https://arxiv.org/abs/1001.0502}{{\ttfamily
  arXiv:1001.0502 [physics.ins-det]}}.

\bibitem{ALICE:pPbMidChdNdEtapaper5020MB}
{\bfseries ALICE} Collaboration, B.~Abelev {\em et~al.}, ``{Pseudorapidity
  density of charged particles in p--Pb collisions at
  $\sqrt{s_{\mathrm{NN}}}~=~5.02$~Te\kern-.1emV\xspace}'',
  \href{https://doi.org/10.1103/PhysRevLett.110.032301}{{\em Phys. Rev. Lett.}
  {\bfseries 110} (2013) 032301},
  \href{https://arxiv.org/abs/1210.3615}{{\ttfamily arXiv:1210.3615
  [nucl-ex]}}.

\bibitem{ALICE:ChPrDefinition}
{\bfseries ALICE} Collaboration, S.~Acharya {\em et~al.}, ``{The ALICE
  definition of primary particles}'', ALICE-PUBLIC-2017-005.
  \url{https://cds.cern.ch/record/2270008}.

\bibitem{ALICE:MBMidrapidity2760}
{\bfseries ALICE} Collaboration, K.~Aamodt {\em et~al.}, ``{Charged-particle
  multiplicity density at midrapidity in central Pb--Pb collisions at
  $\sqrt{s_{\mathrm{NN}}}~=~2.76$~Te\kern-.1emV\xspace}'',
  \href{https://doi.org/10.1103/PhysRevLett.105.252301}{{\em Phys. Rev. Lett.}
  {\bfseries 105} (2010) 252301},
  \href{https://arxiv.org/abs/1011.3916}{{\ttfamily arXiv:1011.3916
  [nucl-ex]}}.

\bibitem{BayesUnfolding}
G.~D'Agostini, ``{A multidimensional unfolding method based on Bayes'
  theorem}'', \href{https://doi.org/10.1016/0168-9002(95)00274-X}{{\em Nuclear
  Instruments and Methods in Physics Research Section A: Accelerators,
  Spectrometers, Detectors and Associated Equipment} {\bfseries 362} (1995)
  487--498}.

\bibitem{roounfold}
T.~Adye, ``{Unfolding algorithms and tests using RooUnfold}'',
  \href{https://arxiv.org/abs/1105.1160}{{\ttfamily arXiv:1105.1160
  [physics.data-an]}}.

\bibitem{Barlow:2003sg}
R.~Barlow, ``{Asymmetric systematic errors}'', MAN-HEP-03-02,
  \href{https://arxiv.org/abs/physics/0306138}{{\ttfamily
  arXiv:physics/0306138}}.

\bibitem{UA5:1988gup}
{\bfseries UA5} Collaboration, R.~E. Ansorge {\em et~al.}, ``{Charged-particle
  multiplicity distributions at 200 GeV and 900 GeV centre-of-mass energy}'',
  \href{https://doi.org/10.1007/BF01506531}{{\em Z. Phys. C} {\bfseries 43}
  (1989) 357}.

\bibitem{Ghosh:2012xh}
P.~Ghosh, ``{Negative binomial multiplicity distribution in proton-proton
  collisions in limited pseudorapidity intervals at LHC up to $\sqrt{s} = 7$
  TeV and the clan model}'',
  \href{https://doi.org/10.1103/PhysRevD.85.054017}{{\em Phys. Rev. D}
  {\bfseries 85} (2012) 054017},
  \href{https://arxiv.org/abs/1202.4221}{{\ttfamily arXiv:1202.4221 [hep-ph]}}.

\bibitem{Giovannini:1998zb}
A.~Giovannini and R.~Ugoccioni, ``{Possible scenarios for soft and semihard
  components structure in central hadron--hadron collisions in the TeV
  region}'', \href{https://doi.org/10.1103/PhysRevD.69.059903}{{\em Phys. Rev.
  D} {\bfseries 59} (1999) 094020},
  \href{https://arxiv.org/abs/hep-ph/9810446}{{\ttfamily
  arXiv:hep-ph/9810446}}. [Erratum: Phys.Rev.D 69, 059903 (2004)].

\bibitem{Giovannini:1999tw}
A.~Giovannini and R.~Ugoccioni, ``{Possible scenarios for soft and semihard
  components structure in central hadron--hadron collisions in the TeV region:
  pseudorapidity intervals}'',
  \href{https://doi.org/10.1103/PhysRevD.60.074027}{{\em Phys. Rev. D}
  {\bfseries 60} (1999) 074027},
  \href{https://arxiv.org/abs/hep-ph/9905210}{{\ttfamily
  arXiv:hep-ph/9905210}}.

\bibitem{Koba:1972ng}
Z.~Koba, H.~B. Nielsen, and P.~Olesen, ``{Scaling of multiplicity distributions
  in high-energy hadron collisions}'',
  \href{https://doi.org/10.1016/0550-3213(72)90551-2}{{\em Nucl. Phys. B}
  {\bfseries 40} (1972) 317--334}.

\bibitem{Schenke:2013dpa}
B.~Schenke, P.~Tribedy, and R.~Venugopalan, ``{Multiplicity distributions in
  pp, p--A and AA collisions from Yang-Mills dynamics}'',
  \href{https://doi.org/10.1103/PhysRevC.89.024901}{{\em Phys. Rev. C}
  {\bfseries 89} (2014) 024901},
  \href{https://arxiv.org/abs/1311.3636}{{\ttfamily arXiv:1311.3636 [hep-ph]}}.

\bibitem{CGCmodel}
E.~Iancu and R.~Venugopalan,
  \href{https://doi.org/10.1142/9789812795533_0005}{{\em {The Color glass
  condensate and high-energy scattering in QCD}}}.
\newblock World Scientific, Singapore, 3, 2003.
\newblock \href{https://arxiv.org/abs/hep-ph/0303204}{{\ttfamily
  arXiv:hep-ph/0303204}}.

\bibitem{PHOBOS:2010eyu}
{\bfseries PHOBOS} Collaboration, B.~Alver {\em et~al.}, ``{Phobos results on
  charged-particle multiplicity and pseudorapidity distributions in Au+Au,
  Cu+Cu, d+Au, and p+p collisions at ultra-relativistic energies}'',
  \href{https://doi.org/10.1103/PhysRevC.83.024913}{{\em Phys. Rev. C}
  {\bfseries 83} (2011) 024913},
  \href{https://arxiv.org/abs/1011.1940}{{\ttfamily arXiv:1011.1940
  [nucl-ex]}}.

\bibitem{ALICE:2013hur}
{\bfseries ALICE} Collaboration, B.~Abelev {\em et~al.}, ``{Centrality
  determination of Pb--Pb collisions at $\sqrt{s_{\mathrm{NN}}}$ = 2.76 TeV
  with ALICE}'', \href{https://doi.org/10.1103/PhysRevC.88.044909}{{\em Phys.
  Rev. C} {\bfseries 88} (2013) 044909},
  \href{https://arxiv.org/abs/1301.4361}{{\ttfamily arXiv:1301.4361
  [nucl-ex]}}.

\bibitem{ALICE:pPbMidChdNdEtapaper5020Cent}
{\bfseries ALICE} Collaboration, J.~Adam {\em et~al.}, ``{Centrality dependence
  of particle production in p--Pb collisions at
  $\sqrt{s_{\mathrm{NN}}}~=~5.02$~Te\kern-.1emV\xspace}'',
  \href{https://doi.org/10.1103/PhysRevC.91.064905}{{\em Phys. Rev. C}
  {\bfseries 91} (2015) 064905},
  \href{https://arxiv.org/abs/1412.6828}{{\ttfamily arXiv:1412.6828
  [nucl-ex]}}.

\bibitem{UA5:1986ye}
{\bfseries UA5} Collaboration, G.~J. Alner {\em et~al.}, ``{Scaling of
  pseudorapidity distributions at c.m. energies up to 0.9 TeV}'',
  \href{https://doi.org/10.1007/BF01410446}{{\em Z. Phys. C} {\bfseries 33}
  (1986) 1--6}.

\bibitem{ALICE:pPbMidChdNdEtapaper8160}
{\bfseries ALICE} Collaboration, S.~Acharya {\em et~al.}, ``{Charged-particle
  pseudorapidity density at midrapidity in p--Pb collisions at
  $\sqrt{s_{\mathrm{NN}}}~=~8.16$~Te\kern-.1emV\xspace}'',
  \href{https://doi.org/10.1140/epjc/s10052-019-6801-9}{{\em Eur. Phys. J. C}
  {\bfseries 79} (2019) 307},
  \href{https://arxiv.org/abs/1812.01312}{{\ttfamily arXiv:1812.01312
  [nucl-ex]}}.

\bibitem{CMS:ChPrMidpPb5TeV}
{\bfseries CMS} Collaboration, A.~M.~Sirunyan {\em et~al.}, ``{Pseudorapidity
  distributions of charged hadrons in p--Pb collisions at
  $\sqrt{s_{\mathrm{NN}}}$ = 5.02 and 8.16~TeV}'',
  \href{https://doi.org/10.1007/JHEP01(2018)045}{{\em JHEP} {\bfseries 01}
  (2018) 045}, \href{https://arxiv.org/abs/1710.09355}{{\ttfamily
  arXiv:1710.09355 [hep-ex]}}.

\end{thebibliography}\endgroup

\newpage
\appendix

%
%

\section{The ALICE Collaboration}
\label{app:collab}
\begin{flushleft} 
\small

S.~Acharya\,\orcidlink{0000-0002-9213-5329}\,$^{\rm 50}$, 
A.~Agarwal$^{\rm 133}$, 
G.~Aglieri Rinella\,\orcidlink{0000-0002-9611-3696}\,$^{\rm 32}$, 
L.~Aglietta\,\orcidlink{0009-0003-0763-6802}\,$^{\rm 24}$, 
M.~Agnello\,\orcidlink{0000-0002-0760-5075}\,$^{\rm 29}$, 
N.~Agrawal\,\orcidlink{0000-0003-0348-9836}\,$^{\rm 25}$, 
Z.~Ahammed\,\orcidlink{0000-0001-5241-7412}\,$^{\rm 133}$, 
S.~Ahmad\,\orcidlink{0000-0003-0497-5705}\,$^{\rm 15}$, 
S.U.~Ahn\,\orcidlink{0000-0001-8847-489X}\,$^{\rm 71}$, 
I.~Ahuja\,\orcidlink{0000-0002-4417-1392}\,$^{\rm 36}$
% ...

\section*{Affiliation Notes}

$^{\rm I}$ Deceased\\
$^{\rm II}$ Also at: Max-Planck-Institut fur Physik, Munich, Germany\\
$^{\rm III}$ Also at: Italian National Agency for New Technologies, Energy and Sustainable Economic Development (ENEA), Bologna, Italy\\
$^{\rm IV}$ Also at: Dipartimento DET del Politecnico di Torino, Turin, Italy\\
$^{\rm V}$ Also at: Department of Applied Physics, Aligarh Muslim University, Aligarh, India\\
$^{\rm VI}$ Also at: Institute of Theoretical Physics, University of Wroclaw, Poland\\
$^{\rm VII}$ Also at: Facultad de Ciencias, Universidad Nacional Aut\'{o}noma de M\'{e}xico, Mexico City, Mexico\\

\section*{Collaboration Institutes}

$^{1}$ A.I. Alikhanyan National Science Laboratory (Yerevan Physics Institute) Foundation, Yerevan, Armenia\\
$^{2}$ AGH University of Krakow, Cracow, Poland\\
$^{3}$ Bogolyubov Institute for Theoretical Physics, National Academy of Sciences of Ukraine, Kiev, Ukraine\\
$^{4}$ Bose Institute, Department of Physics  and Centre for Astroparticle Physics and Space Science (CAPSS), Kolkata, India\\
$^{5}$ California Polytechnic State University, San Luis Obispo, California, United States\\
$^{6}$ Central China Normal University, Wuhan, China\\
$^{7}$ Centro de Aplicaciones Tecnol\'{o}gicas y Desarrollo Nuclear (CEADEN), Havana, Cuba\\
$^{8}$ Centro de Investigaci\'{o}n y de Estudios Avanzados (CINVESTAV), Mexico City and M\'{e}rida, Mexico\\
$^{9}$ Chicago State University, Chicago, Illinois, United States\\
$^{10}$ China Nuclear Data Center, China Institute of Atomic Energy, Beijing, China\\
$^{11}$ China University of Geosciences, Wuhan, China\\
$^{12}$ Chungbuk National University, Cheongju, Republic of Korea\\
$^{13}$ Comenius University Bratislava, Faculty of Mathematics, Physics and Informatics, Bratislava, Slovak Republic\\
$^{14}$ Creighton University, Omaha, Nebraska, United States\\
$^{15}$ Department of Physics, Aligarh Muslim University, Aligarh, India\\
$^{16}$ Department of Physics, Pusan National University, Pusan, Republic of Korea\\
$^{17}$ Department of Physics, Sejong University, Seoul, Republic of Korea\\
$^{18}$ Department of Physics, University of California, Berkeley, California, United States\\
$^{19}$ Department of Physics, University of Oslo, Oslo, Norway\\
$^{20}$ Department of Physics and Technology, University of Bergen, Bergen, Norway\\
$^{21}$ Dipartimento di Fisica, Universit\`{a} di Pavia, Pavia, Italy\\
$^{22}$ Dipartimento di Fisica dell'Universit\`{a} and Sezione INFN, Cagliari, Italy\\
$^{23}$ Dipartimento di Fisica dell'Universit\`{a} and Sezione INFN, Trieste, Italy\\
$^{24}$ Dipartimento di Fisica dell'Universit\`{a} and Sezione INFN, Turin, Italy\\
$^{25}$ Dipartimento di Fisica e Astronomia dell'Universit\`{a} and Sezione INFN, Bologna, Italy\\
$^{26}$ Dipartimento di Fisica e Astronomia dell'Universit\`{a} and Sezione INFN, Catania, Italy\\
$^{27}$ Dipartimento di Fisica e Astronomia dell'Universit\`{a} and Sezione INFN, Padova, Italy\\
$^{28}$ Dipartimento di Fisica `E.R.~Caianiello' dell'Universit\`{a} and Gruppo Collegato INFN, Salerno, Italy\\
$^{29}$ Dipartimento DISAT del Politecnico and Sezione INFN, Turin, Italy\\
$^{30}$ Dipartimento di Scienze MIFT, Universit\`{a} di Messina, Messina, Italy\\
$^{31}$ Dipartimento Interateneo di Fisica `M.~Merlin' and Sezione INFN, Bari, Italy\\
$^{32}$ European Organization for Nuclear Research (CERN), Geneva, Switzerland\\
$^{33}$ Faculty of Electrical Engineering, Mechanical Engineering and Naval Architecture, University of Split, Split, Croatia\\
$^{34}$ Faculty of Nuclear Sciences and Physical Engineering, Czech Technical University in Prague, Prague, Czech Republic\\
$^{35}$ Faculty of Physics, Sofia University, Sofia, Bulgaria\\
$^{36}$ Faculty of Science, P.J.~\v{S}af\'{a}rik University, Ko\v{s}ice, Slovak Republic\\
$^{37}$ Faculty of Technology, Environmental and Social Sciences, Bergen, Norway\\
$^{38}$ Frankfurt Institute for Advanced Studies, Johann Wolfgang Goethe-Universit\"{a}t Frankfurt, Frankfurt, Germany\\
$^{39}$ Fudan University, Shanghai, China\\
$^{40}$ Gangneung-Wonju National University, Gangneung, Republic of Korea\\
$^{41}$ Gauhati University, Department of Physics, Guwahati, India\\
$^{42}$ Helmholtz-Institut f\"{u}r Strahlen- und Kernphysik, Rheinische Friedrich-Wilhelms-Universit\"{a}t Bonn, Bonn, Germany\\
$^{43}$ Helsinki Institute of Physics (HIP), Helsinki, Finland\\
$^{44}$ High Energy Physics Group,  Universidad Aut\'{o}noma de Puebla, Puebla, Mexico\\
$^{45}$ Horia Hulubei National Institute of Physics and Nuclear Engineering, Bucharest, Romania\\
$^{46}$ HUN-REN Wigner Research Centre for Physics, Budapest, Hungary\\
$^{47}$ Indian Institute of Technology Bombay (IIT), Mumbai, India\\
$^{48}$ Indian Institute of Technology Indore, Indore, India\\
$^{49}$ INFN, Laboratori Nazionali di Frascati, Frascati, Italy\\
$^{50}$ INFN, Sezione di Bari, Bari, Italy\\
$^{51}$ INFN, Sezione di Bologna, Bologna, Italy\\
$^{52}$ INFN, Sezione di Cagliari, Cagliari, Italy\\
$^{53}$ INFN, Sezione di Catania, Catania, Italy\\
$^{54}$ INFN, Sezione di Padova, Padova, Italy\\
$^{55}$ INFN, Sezione di Pavia, Pavia, Italy\\
$^{56}$ INFN, Sezione di Torino, Turin, Italy\\
$^{57}$ INFN, Sezione di Trieste, Trieste, Italy\\
$^{58}$ Inha University, Incheon, Republic of Korea\\
$^{59}$ Institute for Gravitational and Subatomic Physics (GRASP), Utrecht University/Nikhef, Utrecht, Netherlands\\
$^{60}$ Institute of Experimental Physics, Slovak Academy of Sciences, Ko\v{s}ice, Slovak Republic\\
$^{61}$ Institute of Physics, Homi Bhabha National Institute, Bhubaneswar, India\\
$^{62}$ Institute of Physics of the Czech Academy of Sciences, Prague, Czech Republic\\
$^{63}$ Institute of Space Science (ISS), Bucharest, Romania\\
$^{64}$ Institut f\"{u}r Kernphysik, Johann Wolfgang Goethe-Universit\"{a}t Frankfurt, Frankfurt, Germany\\
$^{65}$ Instituto de Ciencias Nucleares, Universidad Nacional Aut\'{o}noma de M\'{e}xico, Mexico City, Mexico\\
$^{66}$ Instituto de F\'{i}sica, Universidade Federal do Rio Grande do Sul (UFRGS), Porto Alegre, Brazil\\
$^{67}$ Instituto de F\'{\i}sica, Universidad Nacional Aut\'{o}noma de M\'{e}xico, Mexico City, Mexico\\
$^{68}$ iThemba LABS, National Research Foundation, Somerset West, South Africa\\
$^{69}$ Jeonbuk National University, Jeonju, Republic of Korea\\
$^{70}$ Johann-Wolfgang-Goethe Universit\"{a}t Frankfurt Institut f\"{u}r Informatik, Fachbereich Informatik und Mathematik, Frankfurt, Germany\\
$^{71}$ Korea Institute of Science and Technology Information, Daejeon, Republic of Korea\\
$^{72}$ Laboratoire de Physique Subatomique et de Cosmologie, Universit\'{e} Grenoble-Alpes, CNRS-IN2P3, Grenoble, France\\
$^{73}$ Lawrence Berkeley National Laboratory, Berkeley, California, United States\\
$^{74}$ Lund University Department of Physics, Division of Particle Physics, Lund, Sweden\\
$^{75}$ Nagasaki Institute of Applied Science, Nagasaki, Japan\\
$^{76}$ Nara Women{'}s University (NWU), Nara, Japan\\
$^{77}$ National and Kapodistrian University of Athens, School of Science, Department of Physics , Athens, Greece\\
$^{78}$ National Centre for Nuclear Research, Warsaw, Poland\\
$^{79}$ National Institute of Science Education and Research, Homi Bhabha National Institute, Jatni, India\\
$^{80}$ National Nuclear Research Center, Baku, Azerbaijan\\
$^{81}$ National Research and Innovation Agency - BRIN, Jakarta, Indonesia\\
$^{82}$ Niels Bohr Institute, University of Copenhagen, Copenhagen, Denmark\\
$^{83}$ Nikhef, National institute for subatomic physics, Amsterdam, Netherlands\\
$^{84}$ Nuclear Physics Group, STFC Daresbury Laboratory, Daresbury, United Kingdom\\
$^{85}$ Nuclear Physics Institute of the Czech Academy of Sciences, Husinec-\v{R}e\v{z}, Czech Republic\\
$^{86}$ Oak Ridge National Laboratory, Oak Ridge, Tennessee, United States\\
$^{87}$ Ohio State University, Columbus, Ohio, United States\\
$^{88}$ Physics department, Faculty of science, University of Zagreb, Zagreb, Croatia\\
$^{89}$ Physics Department, Panjab University, Chandigarh, India\\
$^{90}$ Physics Department, University of Jammu, Jammu, India\\
$^{91}$ Physics Program and International Institute for Sustainability with Knotted Chiral Meta Matter (WPI-SKCM$^{2}$), Hiroshima University, Hiroshima, Japan\\
$^{92}$ Physikalisches Institut, Eberhard-Karls-Universit\"{a}t T\"{u}bingen, T\"{u}bingen, Germany\\
$^{93}$ Physikalisches Institut, Ruprecht-Karls-Universit\"{a}t Heidelberg, Heidelberg, Germany\\
$^{94}$ Physik Department, Technische Universit\"{a}t M\"{u}nchen, Munich, Germany\\
$^{95}$ Politecnico di Bari and Sezione INFN, Bari, Italy\\
$^{96}$ Research Division and ExtreMe Matter Institute EMMI, GSI Helmholtzzentrum f\"ur Schwerionenforschung GmbH, Darmstadt, Germany\\
$^{97}$ Saga University, Saga, Japan\\
$^{98}$ Saha Institute of Nuclear Physics, Homi Bhabha National Institute, Kolkata, India\\
$^{99}$ School of Physics and Astronomy, University of Birmingham, Birmingham, United Kingdom\\
$^{100}$ Secci\'{o}n F\'{\i}sica, Departamento de Ciencias, Pontificia Universidad Cat\'{o}lica del Per\'{u}, Lima, Peru\\
$^{101}$ Stefan Meyer Institut f\"{u}r Subatomare Physik (SMI), Vienna, Austria\\
$^{102}$ SUBATECH, IMT Atlantique, Nantes Universit\'{e}, CNRS-IN2P3, Nantes, France\\
$^{103}$ Sungkyunkwan University, Suwon City, Republic of Korea\\
$^{104}$ Suranaree University of Technology, Nakhon Ratchasima, Thailand\\
$^{105}$ Technical University of Ko\v{s}ice, Ko\v{s}ice, Slovak Republic\\
$^{106}$ The Henryk Niewodniczanski Institute of Nuclear Physics, Polish Academy of Sciences, Cracow, Poland\\
$^{107}$ The University of Texas at Austin, Austin, Texas, United States\\
$^{108}$ Universidad Aut\'{o}noma de Sinaloa, Culiac\'{a}n, Mexico\\
$^{109}$ Universidade de S\~{a}o Paulo (USP), S\~{a}o Paulo, Brazil\\
$^{110}$ Universidade Estadual de Campinas (UNICAMP), Campinas, Brazil\\
$^{111}$ Universidade Federal do ABC, Santo Andre, Brazil\\
$^{112}$ Universitatea Nationala de Stiinta si Tehnologie Politehnica Bucuresti, Bucharest, Romania\\
$^{113}$ University of Derby, Derby, United Kingdom\\
$^{114}$ University of Houston, Houston, Texas, United States\\
$^{115}$ University of Jyv\"{a}skyl\"{a}, Jyv\"{a}skyl\"{a}, Finland\\
$^{116}$ University of Kansas, Lawrence, Kansas, United States\\
$^{117}$ University of Liverpool, Liverpool, United Kingdom\\
$^{118}$ University of Science and Technology of China, Hefei, China\\
$^{119}$ University of South-Eastern Norway, Kongsberg, Norway\\
$^{120}$ University of Tennessee, Knoxville, Tennessee, United States\\
$^{121}$ University of the Witwatersrand, Johannesburg, South Africa\\
$^{122}$ University of Tokyo, Tokyo, Japan\\
$^{123}$ University of Tsukuba, Tsukuba, Japan\\
$^{124}$ Universit\"{a}t M\"{u}nster, Institut f\"{u}r Kernphysik, M\"{u}nster, Germany\\
$^{125}$ Universit\'{e} Clermont Auvergne, CNRS/IN2P3, LPC, Clermont-Ferrand, France\\
$^{126}$ Universit\'{e} de Lyon, CNRS/IN2P3, Institut de Physique des 2 Infinis de Lyon, Lyon, France\\
$^{127}$ Universit\'{e} de Strasbourg, CNRS, IPHC UMR 7178, F-67000 Strasbourg, France, Strasbourg, France\\
$^{128}$ Universit\'{e} Paris-Saclay, Centre d'Etudes de Saclay (CEA), IRFU, D\'{e}partment de Physique Nucl\'{e}aire (DPhN), Saclay, France\\
$^{129}$ Universit\'{e}  Paris-Saclay, CNRS/IN2P3, IJCLab, Orsay, France\\
$^{130}$ Universit\`{a} degli Studi di Foggia, Foggia, Italy\\
$^{131}$ Universit\`{a} del Piemonte Orientale, Vercelli, Italy\\
$^{132}$ Universit\`{a} di Brescia, Brescia, Italy\\
$^{133}$ Variable Energy Cyclotron Centre, Homi Bhabha National Institute, Kolkata, India\\
$^{134}$ Warsaw University of Technology, Warsaw, Poland\\
$^{135}$ Wayne State University, Detroit, Michigan, United States\\
$^{136}$ Yale University, New Haven, Connecticut, United States\\
$^{137}$ Yildiz Technical University, Istanbul, Turkey\\
$^{138}$ Yonsei University, Seoul, Republic of Korea\\
$^{139}$ Affiliated with an institute formerly covered by a cooperation agreement with CERN\\
$^{140}$ Affiliated with an international laboratory covered by a cooperation agreement with CERN.\\

\end{flushleft} 
  
\end{document}